\documentclass[12pt]{article}
\usepackage{epsfig}
\usepackage{graphicx}
\usepackage{amsmath}
\usepackage{slashed}
\newcommand{\beq}{\begin{equation}}
\newcommand{\eeq}{\end{equation}}
\newcommand{\bea}{\begin{eqnarray}}
\newcommand{\eea}{\end{eqnarray}}
\newcommand{\bem}{\begin{multline}}
\newcommand{\eem}{\end{multline}}
\newcommand{\beg}{\begin{gather}}
\newcommand{\eeg}{\end{gather}}
\newcommand{\noi}{\noindent}

\newcommand{\uk}{\underline{k}}
\newcommand{\ux}{\underline{x}}
\newcommand{\uy}{\underline{y}}
\newcommand{\uz}{\underline{z}}
\newcommand{\ub}{\underline{b}}
\newcommand{\s}{\slashed}
\newcommand{\lt}{<}
\newcommand{\gt}{>}
\def\ud{\underline}

\newcommand{\as}{\alpha_s}

\newcommand{\bas}{{\bar\alpha}_s}
\def\eq#1{{Eq.~(\ref{#1})}}
\def\fig#1{{Fig.~\ref{#1}}}
\newcommand{\ben}{\begin{eqnarray*}}
\newcommand{\een}{\end{eqnarray*}}
\newcommand{\un}[1]{\underline{#1}}

\begin{document}
\title{{\bf Baryon Stopping in Proton-Nucleus Collisions
\\[1.5cm] }}
\author{
{\bf Javier
  L. Albacete\thanks{e-mail: albacete@mps.ohio-state.edu}\hspace{0.2cm} and 
 Yuri V.\ Kovchegov\thanks{e-mail: yuri@mps.ohio-state.edu}}
\\[1cm] {\it\small Department of Physics, The Ohio State University}\\ 
{\it\small Columbus, OH 43210,USA}\\[5mm]}

\date{May 2006}

\maketitle

\thispagestyle{empty}

\begin{abstract}
We calculate the inclusive small-$x$ valence quark production cross
section in proton-nucleus collisions at high energies. The calculation
is performed in the framework of the Color Glass Condensate
formalism. We consider both the case when the valence quark originates
inside the nucleus and the case when it originates inside the
proton. We first calculate the cross section in the quasi-classical
approximation resumming the multiple rescatterings with the
nucleus. Then we include the the effects of double logarithmic reggeon
evolution and leading logarithmic gluon evolution in the obtained
cross section. The calculated nuclear modification factor for the
stopped baryons exhibits Cronin enhancement in the quasi-classical
approximation and suppression at high energies/rapidities when quantum
evolution corrections are included, providing a new observable which
can be used to test Color Glass physics.
\end{abstract}

\thispagestyle{empty}

\newpage

\setcounter{page}{1}


\section{Introduction}

One of the most striking features of experimental data in nuclear
collisions at high energies is the large stopping suffered by the
nucleons entering the collision \cite{NA49,NA35,BRAHMS}. This stopping
translates into a net baryon number transfer from the forward to the
central rapidity region, and has already been observed in
proton-nucleus and nucleus-nucleus collisions at CERN \cite{NA49,NA35}
and in $Au+Au$ collisions at RHIC \cite{BRAHMS}.  The description of
this phenomenon has been previously approached in the framework of
non-perturbative models for hadron collisions
\cite{Veneziano,RV,KZ,Dima96,CS}. In \cite{Veneziano,RV,KZ,Dima96,CS}
the baryon number transfer along a large rapidity gap is attributed to
topological excitations of the gluon field.

In this work, following \cite{IKMT}, we consider the problem of baryon
stopping from a purely perturbative perspective in which the carriers
of baryon number are the energetic valence quarks entering the
collision, that loose much of their longitudinal momenta in the
collision and are driven towards the central rapidity region via hard
gluon emissions. Valence quark production in the scattering of a
proton on a nucleus is calculated in the framework of the Color
Glass Condensate (CGC) formalism
\cite{glr,mv,yuri1,jkmw,dip,yuri,bal,JKLW,FILM}. The effects of strong 
gluonic fields in the small-$x$ hadronic or nuclear wave functions on
baryon number transfer were first considered by one of the authors
with collaborators in \cite{IKMT}. The valence quark distribution at
small Bjorken $x$ was first studied in \cite{IKMT} in the
quasi-classical approximation of McLerran-Venugopalan model \cite{mv}.
Quantum evolution corrections come into the obtained valence quark
distribution through a non-linear evolution equation which was derived
in \cite{IKMT}. The linear part of the evolution equation from
\cite{IKMT} corresponds to the well-known double logarithmic reggeon
evolution equation from \cite{KL,Kirschner1,Kirschner2,GR}. It is an
interesting property of quark evolution that at the leading order each
power of the coupling constant $\as$ is accompanied by {\sl two}
powers of the logarithm of center-of-mass energy $\ln s$, or,
equivalently, of the logarithm $\ln 1/x$ of inverse Bjorken $x$
\cite{KL,Kirschner1,Kirschner2,GR}. The resummation parameter is thus 
$\as \, \ln^2 1/x$, unlike, say, the BFKL equation \cite{BFKL}, which
resums powers of $\as \, \ln 1/x$. Due to that property one may expect
the small-$x$ evolution corrections to manifest themselves earlier
(i.e., at larger values of $x$ and smaller rapidities) in observables
related to valence quarks. The non-linear part of the evolution
equation derived in \cite{IKMT} is due to gluon evolution effects,
similar to those resummed in \cite{yuri,bal,JKLW,FILM}. While the
linear reggeon evolution \cite{KL,Kirschner1,Kirschner2,GR} tends to
increase the small-$x$ valence quark distribution, the effect of
nonlinear evolution equation \cite{IKMT} is to reduce the valence
quark distribution at very small-$x$ as compared to the linear
evolution predictions.

Here we continue the work done in \cite{IKMT} by calculating the
inclusive production cross section for small-$x$ valence quarks in
$pA$ collisions. We perform the calculations both in the
quasi-classical approximation \cite{mv,yuri1} and including quantum
evolution from \cite{IKMT}. We use the techniques for calculation of
inclusive cross sections developed in \cite{KM,KT} (for a review see
\cite{JMK}). The calculation of inclusive gluon production in pA
collisions was performed in \cite{KM,DM} in the quasi-classical
approximation and in \cite{KT} the effects of quantum evolution were
included in the cross section. Our calculations will proceed along
similar lines. In Section 2 we derive the expression for valence quark
production in the quasi-classical approximation resumming multiple
rescatterings of the incoming proton and produced valence quark on the
nucleons in the target nucleus. Here one has to separately consider
two cases: the valence quark may originate either in the proton or in
the nucleus. The stopping of valence quarks in these two cases is
clearly distinguishable experimentally: the rapidity distribution of
stopped valence quarks from the proton scales as $\sim e^{y-Y}$ while
the valence quarks from the nucleus are distributed as $\sim
e^{-y}$. (Here the nucleus has rapidity $0$, the proton has rapidity
$Y$ and the quark is produced at rapidity $y$.) We calculate both
contributions in Sect 2.1 and study their properties in Sect 2.2. Our
results here are complimentary to the valence quark production
calculated in \cite{DJ}: the authors of \cite{DJ} considered
production of a hard valence quark which experiences no recoil and is
produced in the fragmentation region, while here we are interested in
production of a soft valence quark far away (in rapidity) from the
fragmentation region.

As was shown in
\cite{KKT,AAKSW,BKW,KTS,KNST,ktbroadening1,ktbroadening2,Vitev03,ktbroadening3} 
for gluons, multiple rescatterings in the nucleus lead to enhancement
of high-$p_T$ particle production which is similar to the Cronin
effect observed in \cite{Cronin}. Analyzing the nuclear modification
factor resulting from the derived quasi-classical cross section for
valence quark production, we observe that valence quarks also exhibit
Cronin enhancement, as demonstrated in \fig{cronin}.

We continue in Section 3 by including the effects of non-linear
reggeon evolution from \cite{IKMT} into the valence quark production
cross section. The cases of valence quark originating in the proton
and in the nucleus are considered separately in Sections 3.1 and
3.3. Similar to the case of gluon production \cite{KT}, the evolution
between the produced quark and the projectile turns out to be linear
and is given by the reggeon evolution equation \cite{KL} if the quark
comes from the proton and by the BFKL equation \cite{BFKL} if the
quark comes from the nucleus. The evolution between the produced
valence quark and the target nucleus is always non-linear.

As was originally observed in \cite{KLM,KKT,AAKSW} for gluon
production, small-$x$ evolution introduces suppression in the nuclear
modification factor at high energies/rapidities, turning Cronin
enhancement into suppression. These predictions were confirmed by the
forward rapidity particle production data produced by RHIC $d+Au$
experiments \cite{brahms-1,brahms-2,phobosdA,phenixdA,stardA}. In
Section 3.2 we study the effects of non-linear reggeon evolution on
valence quark nuclear modification factor for the case when the
valence quarks are produced from the proton. We observe suppression of
nuclear modification factor which increases with rapidity/energy
similar to the case of the gluons, as shown in \fig{evcron}. Therefore, if
baryon stopping is dominated by perturbative mechanisms, we expect the
nuclear modification factor for net baryons to be suppressed at
forward rapidities in $d+Au$ collisions at RHIC. Thus the nuclear
modification factor for stopped baryons may serve as a new test of
Color Glass Condensate (CGC) dynamics in nuclear collisions, or, at
least it may help determine the relative contributions of perturbative
and non-perturbative baryon stopping.

We conclude in Section 4 by discussing the applicability of obtained
results and the prospects for experimental testing of the derived
formulas.  In order to compare our results to the actual experimental
data produced at RHIC and to make predictions for the LHC, as was done
in \cite{KKT2,JamalH,DHJ1,DGS}, one should implement the fragmentation
mechanism which converts valence quarks into hadrons. This goes beyond
the scope of this paper and is left for future work.


\section{Valence Quark Production in the Quasi-Classical Approximation}

\subsection{Derivation of the Expression for the Production Cross Section}

We are interested in calculating the soft valence quark spectrum in
proton-nucleus collisions at high energies. In what follows we will
assume that the proton is moving ultrarelativistically in the
light-cone 'plus' direction and, consequently, has a large $p_1^+$
momentum component, whereas a nucleon in the nucleus is moving in the
light-cone 'minus' direction with a large $p_2^-$ momentum
component. We employ Sudakov parameterization for the light-cone
components of the momentum of the the produced valence quark:
$k=(\alpha \, p_1^+, \beta \, p_2^-,\uk)$. In the center of mass frame
of the collision, the relation of the dimensionless Sudakov parameters
$\alpha$ and $\beta$ to the rapidity of the produced quark $y$ and the
total rapidity of the collision $Y$ is
\beq
\alpha \, =  \, \frac{|\uk|}{\sqrt{s}}e^{y-Y/2},\qquad 
\beta \, = \, \frac{|\uk|}{\sqrt{s}}e^{-y+Y/2}\,,
\label{alpha}
\eeq
\noi where $s = 2p_1^+ \, p_2^-$ is the square of the center-of-mass energy of the 
collision and the proton has rapidity $y=Y$ while the target nucleus
has $y=0$. The calculation below will be performed in the light-cone
gauge of the proton, $A^+=0$, which for the nucleus moving in the
'minus' direction is equivalent to the covariant gauge, $\partial\cdot
A=0$. This choice completely determines the set of diagrams that are
relevant for our calculation as well as their physical interpretation:
As it was discussed in detail in \cite{KM,Kovchegov:2000hz}, in this
gauge the classical current associated with the nucleus, $J^-$,
remains unchanged throughout the collision and the nuclear effects
become manifest in the form of multiple scatterings between the
nucleons in the nucleus and the rest of the system.

As usual at high energies, the resummation of the multiple
rescatterings is performed in the eikonal approximation, in which the
energetic projectiles (quark or gluon) remain at fixed transverse
coordinate as they propagate through nuclear matter. Hence the
calculation is done in coordinate space. Besides, we restrict the
interactions with the nucleons in the nucleus to two-gluon exchanges,
in the spirit of the quasi-classical approximation. More technically
this implies that the resummation parameter is $\alpha_s^2A^{1/3}$,
with $A$ the atomic number of the nucleus.  We are assuming that the
energy of the collision is large enough for an eikonal description of
the multiple rescatterings to be appropriate, but not large enough for
quantum corrections to become important. Under these two
approximations (eikonal limit and quasi-classical description) the
effects of multiple rescatterings, and therefore all the nuclear
effects, can be recast in the form of Glauber-Mueller
propagators \cite{KM,GM}.

We divide the cross section for soft quark production into two terms:
the first contribution, labeled (a), corresponds to produced valence
quark coming from the proton, while the second one, labeled (b),
corresponds to valence quark coming from one of the nucleons in the
nucleus via interaction with a soft gluon in the proton wavefunction
(see below):
\beq
\frac{d\sigma}{d^2 k \, dy}=\frac{d\sigma^{a}}{d^2 k \, dy} + 
\frac{d\sigma^{b}}{d^2 k \, dy}.
\label{ds}
\eeq

Importantly, the formation time associated with the soft valence quark
production mechanism is, in both cases, much longer than the typical
time scale for the nucleus-proton interactions, which can be
considered instantaneous. Moreover, the diagrams in which the soft
valence quark emission happens during the interaction with the nucleus
are suppressed by powers of the total energy of the collision and,
therefore, can be neglected in the high energy limit. This allows us
to deal with time-ordered diagrams or, more precisely, the calculation
is more convenient (and will be done) in the framework of time-ordered
light-cone perturbation theory \cite{BL}.

\begin{figure}[ht]
\begin{center}
\includegraphics[height=9cm,width=15.5cm]{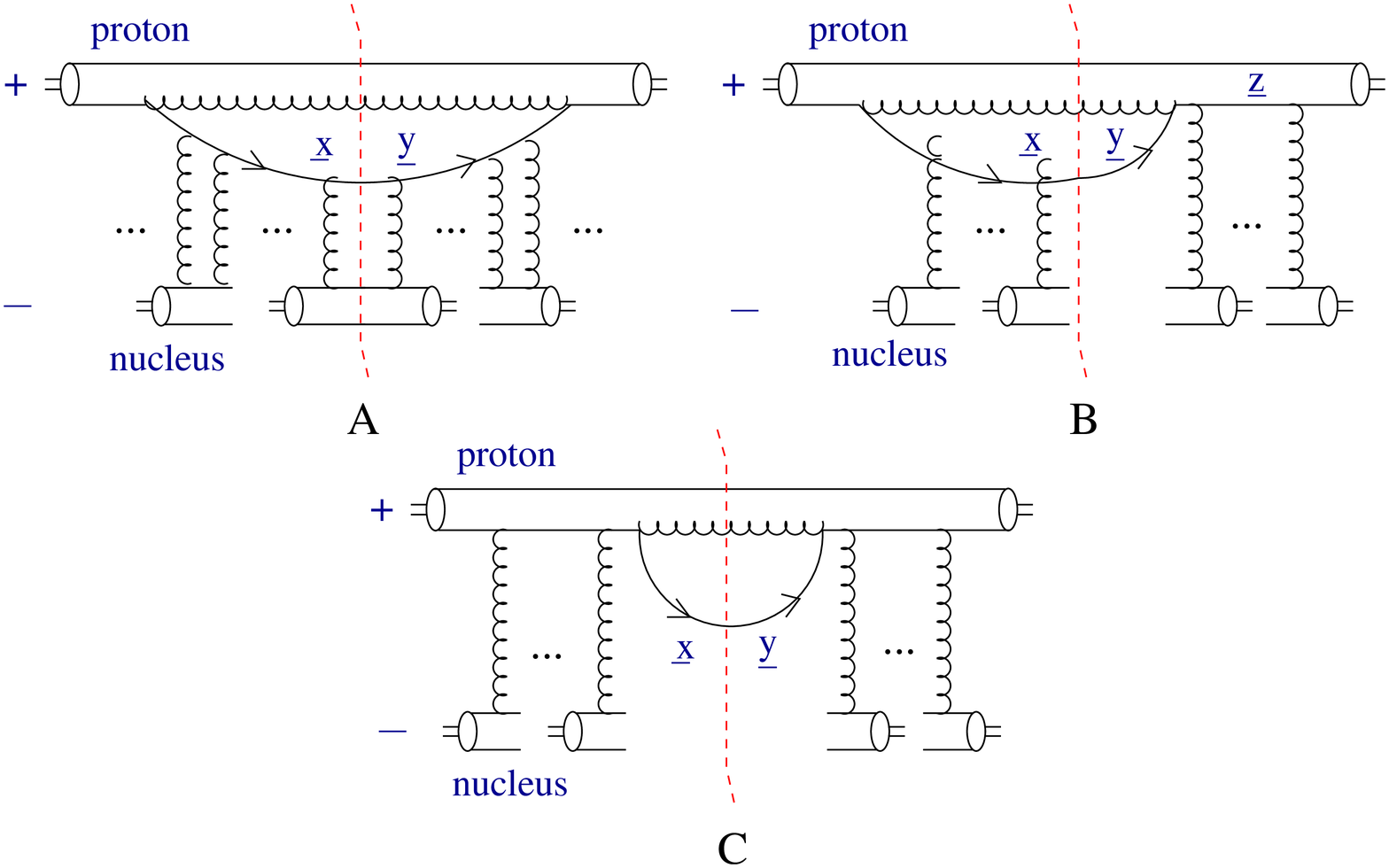}
\end{center}
\caption{Diagrams corresponding to the first contribution to the cross 
section as considered in the text, where the produced valence quark
originates in the proton.}
\label{f1}
\end{figure}
 
First we are going to find the contribution to the cross section in
which the produced soft valence quark originates in the proton
wavefunction. The contributing diagrams are shown in \fig{f1}. The
discussion here will proceed along the lines of the calculation done
in \cite{KM} of the gluon production in pA collisions. The physical
picture is the following: one of the fast valence quarks in the proton
has to emit a hard gluon in order to loose most of its large
longitudinal momentum and, therefore, become soft. This splitting is
given by the soft-quark wavefunction of the proton at the lowest order
in perturbation theory (see appendix A):
\beq
\psi^a_{\sigma,\lambda}(\ux,\uz,\alpha)
=gt^a[1-\sigma\lambda] \, \frac{i}{2\pi} \,
\frac{(\ux-\uz)\cdot\ud{\epsilon}^{\lambda}}{(\ux-\uz)^2}\, ,
\label{qwf}
\eeq
\noi where $\uz$ is the transverse coordinate of the hard gluon (which 
is equal to the transverse coordinate of the fast quark), $\un x$ is
the transverse coordinate of the soft quark, $\sigma$ is the helicity
of the fast quark, $a$ and $\lambda$ are the color and polarization
of the hard gluon. The helicity of the soft quark is equal to the one
of the fast one due to helicity conservation for massless quarks.

Analogously to what was done in \cite{IKMT}, it is convenient to
introduce the soft quark distribution of a proton. In order to do so
we multiply the above wavefunction in \eq{qwf} by its complex
conjugate evaluated at transverse position $\uy$, average over the
helicity of the initial quark and sum over the polarization and color
of the hard gluon, getting
\beq
\frac{dn^q}{d^2 z \, dy} \, = \, \frac{1}{4\pi} \, \frac{\alpha}{2} \, 
\sum_{\sigma,\lambda,a}
\langle\psi^a_{\sigma,\lambda}(\ux,\uz, \alpha)\psi^{*a}_{\sigma,\lambda}
(\uy,\uz, \alpha)
\rangle =\frac{\bar{\alpha}_s \, \alpha \, N_c}{2\pi} \, 
\frac{(\ux-\uz)\cdot(\uy-\uz)}{(\ux-\uz)^2(\uy-\uz)^2}\,,
\label{qdf}
\eeq
where $\bar{\alpha}_s=\alpha_sC_F/\pi$. The factor of $N_c$ on the
right hand side of \eq{qdf} accounts for the three valence quarks in a
proton. All the dynamics corresponding to the soft-quark emission in
the diagrams we are going to calculate is included in the distribution
in \eq{qdf}. 

Our next step is to resum the multiple rescatterings of the
quark-gluon system in the nucleus. The diagram shown in \fig{f1}A
corresponds to the case in which the incoming proton has already
developed a soft valence quark component in its wavefunction before
interacting with the nucleus both in the amplitude and in its complex
conjugate. As is explicitly shown in \fig{f1}A, in this case only the
interactions between the nucleons and the soft-quark line survive. The
interactions between the nucleons and the hard gluon vanish due to
real-virtual cancellation \cite{KM}. This can be understood in the
following terms: since we want to keep the momentum of the produced
soft valence quark fixed, in the transverse coordinate space that we
are going to perform our calculation in, its transverse coordinate in
the amplitude $\ux$ has to be different from the one in the complex
conjugate amplitude $\uy$ \cite{KM,JMK}. The transverse momenta of all
other particles in the final state are integrated over (inclusive
production) making their coordinates in the amplitude and in the
complex conjugate amplitude equal to each other. Therefore, one can
freely move the $t$-channel gluon lines connecting the nucleons with
the hard gluon across the cut without changing either the color factor
or the momentum of the produced quark, but picking up a relative minus
sign in the way, which provides the cancellation.

To calculate the surviving contribution one can proceed either by
direct evaluation of the diagrams or, alternatively, by making use of
the crossing symmetry property \cite{Mueller_cross,JMK} in the
following way: mirror-reflecting the soft quark line in the complex
conjugate amplitude of \fig{f1}A with respect to the final state cut
one is left with the diagram corresponding to the amplitude of a
quark-antiquark dipole of transverse size $|\ux-\uy|$ multiply
scattering off a nucleus. The amplitude for this process is \cite{GM}
\beq
N(\ux,\uy) \, = \, 1 - e^{-|\ux-\uy|^2Q_{sq}^2\ln(1/|\ux-\uy|\Lambda)/4},
\label{ei1}
\eeq
\noi where 
\beq 
Q_{sq}^2=4\pi \, {\alpha}_s^2 \, \frac{C_F}{N_c} \, \rho \, T_A(\ud{b})
\label{qss}
\eeq
\noi is the saturation scale of a quark dipole, $\rho$ is the nuclear density
(normalized to $A$) and $T_A(\ud{b})$ is the nuclear thickness at
impact parameter $\ub$ in its own rest frame, which for a spherical
nucleus of radius $R$ is equal to $T_A(\ud{b})=2\sqrt{R^2-\ud{b}^2}$.

\begin{figure}
\begin{center}
\includegraphics[height=4.5cm,width=9cm]{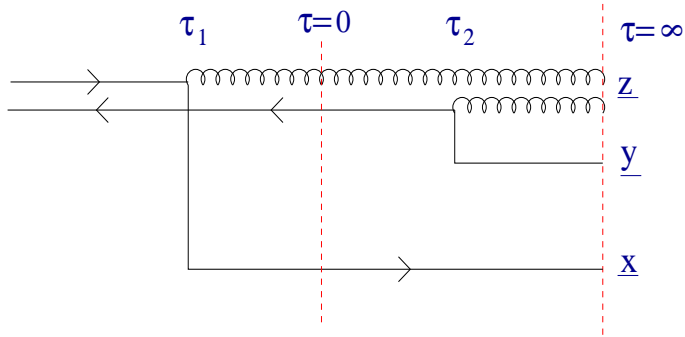}
\end{center}
\caption{Mirror reflection of the diagram in \fig{f1}B. $\tau_{1(2)}$ is the
  emission time in the amplitude (complex conjugate amplitude). The
  interaction time is $\tau=0$, represented in the figure by a
  vertical line and the final state cut is denoted by $\tau = +
  \infty$.}
\label{m1}
\end{figure}

The diagram in which the soft quark is emitted before the interaction
with the nucleus in the amplitude, but after the interaction in the
complex conjugate one is shown in \fig{f1}B.  The interactions between
the nucleus and the hard gluon line in the amplitude do not cancel in
this case since, to move the gluon line attached to the fast quark in
the complex conjugate amplitude across the cut, one is forced to
'jump' over the emission vertex, which would modify the momentum of
the produced quark, ruining the cancellation.  To explicitly include
both the interactions with the $s$-channel quark and gluon lines, we
do not connect the $t$-channel gluon lines to any particular line in
\fig{f1}B. Again, mirror-reflecting the complex conjugate amplitude
with respect to the cut generates the configuration represented in
\fig{m1}. There, the fast valence quark line in the complex conjugate
amplitude turns into an antiquark line overlapping the hard gluon line
in the amplitude at transverse coordinate $\un 0$. Since the whole
system is color neutral, the zero-size antiquark-gluon system has to
be in the $\bar{3}$ representation of $SU(3)$, which together with the
quark line located at the $\ux$ coordinate is equivalent to a quark
dipole multiply rescattering off a nucleus. Therefore, the factor
entering the production cross-section due to interactions in \fig{f1}B
is
\beq
N(\ux,\uz)=1-e^{-|\ux-\uz|^2Q_{sq}^2\ln(1/|\ux-\uz|\Lambda)/4}.
\label{ei2}
\eeq

The contribution of the diagram complex conjugate to \fig{f1}B (not
shown) is calculated in an analogous way yielding
\beq
N(\uy,\uz)=1-e^{-|\uy-\uz|^2Q_{sq}^2\ln(1/|\uy-\uz|\Lambda)/4}.
\label{ei3}
\eeq
\vspace{0.5cm}

The diagram in \fig{f1}C gives a zero contribution to the production
cross-section, since, due to real-virtual cancellations, all the
interactions with the target cancel, leading to $N=0$.

Finally, adding up the contributions to the production cross sections
of the diagrams in \fig{f1} gives
\begin{gather}
\frac{d\sigma^a}{d^2k\,dy}=\frac{1}{(2\pi)^2} \, \int\! d^2 x\,d^2 y\,d^2 z\, 
e^{i\uk(\ux-\uy)} \, \frac{dn^q}{d^2 z\,dy} \nonumber\\
\times \, \left[e^{-(\ux-\uy)^2Q_{sq}^2\ln(1/|\ux-\uy|\Lambda)/4}-
e^{-(\ux-\uz)^2Q_{sq}^2\ln(1/|\ux-\uz|\Lambda)/4}-
e^{-(\uy-\uz)^2Q_{sq}^2\ln(1/|\uy-\uz|\Lambda)/4}+1\right].
\label{dsa}
\end{gather}

\begin{figure}[ht]
\begin{center}
\includegraphics[height=6cm,width=11.5cm]{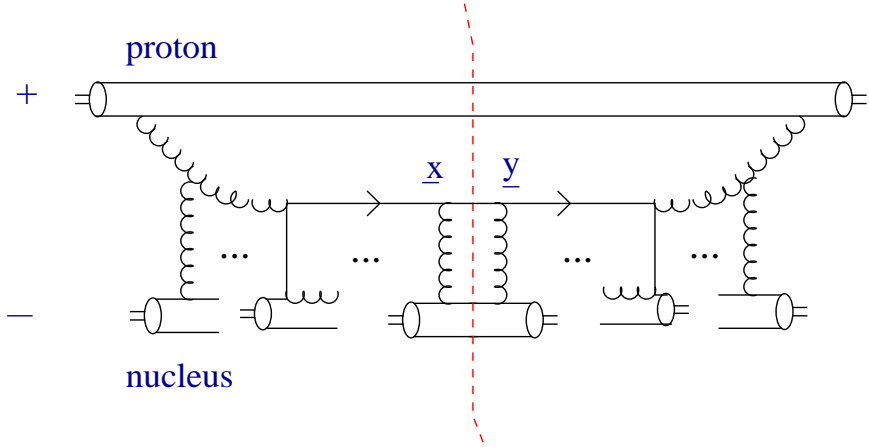}
\end{center}
\caption{Valence quark production diagram corresponding to the 
second case considered in the text, where the produced soft valence
quark originates from the nuclear wave function.}
\label{f2}
\end{figure}

Our next step is to calculate the diagram in which the produced
valence quark originates in the nucleus wavefunction, an example of
which is shown in \fig{f2}. Before starting this calculation, let us
discuss the diagrams that do not contribute to the cross-section in
this case.  First, the emission of the hard gluon cannot happen via
interaction with any other nucleon in the nucleus. In its rest frame
the nucleus is a dilute system, i.e, its constituents nucleons are
spatially separated from each other. In a frame in which the nucleus
is fast moving in the light-cone 'minus' direction, diluteness of the
nucleus translates into the fact that the nucleons are strongly
localized around fixed well differentiated $x^+$ coordinates:
$x_1^+\lt x_2^+\dots\lt x_A^+$. Moreover, the gluon field of the
nucleus as given by the classical equations of motion in the $A^+=0$
gauge is equivalent to the gluon field in covariant gauge and is
strongly localized around the $x^+$ coordinates of the nucleons
\cite{GM,yuri1}: $\ud{A}\sim\Sigma_i \delta(x^+ - x^+_i)$. Therefore, the 
interactions between the nucleons are suppressed by powers of $A$. In
the leading order in $\as^2 A^{1/3}$ calculation, corresponding to the
quasi-classical limit used here, the nucleons can be assumed as not
interacting with each other. On the other hand, those diagrams in
which a nucleon exchanges one gluon with one of the quarks in the
proton and subsequently emits a hard gluon are suppressed by a power
of center-of-mass energy at high energies in $A^+=0$ light-cone gauge
and are also subleading. This leaves us with a single possibility for
the nucleon's valence quark to be produced at central rapidity: it may
only happen via interaction with a soft gluon in the proton's
wavefunction, as shown in \fig{f2}.

\begin{figure}[ht]
\begin{center}
\includegraphics[height=6cm,width=7cm]{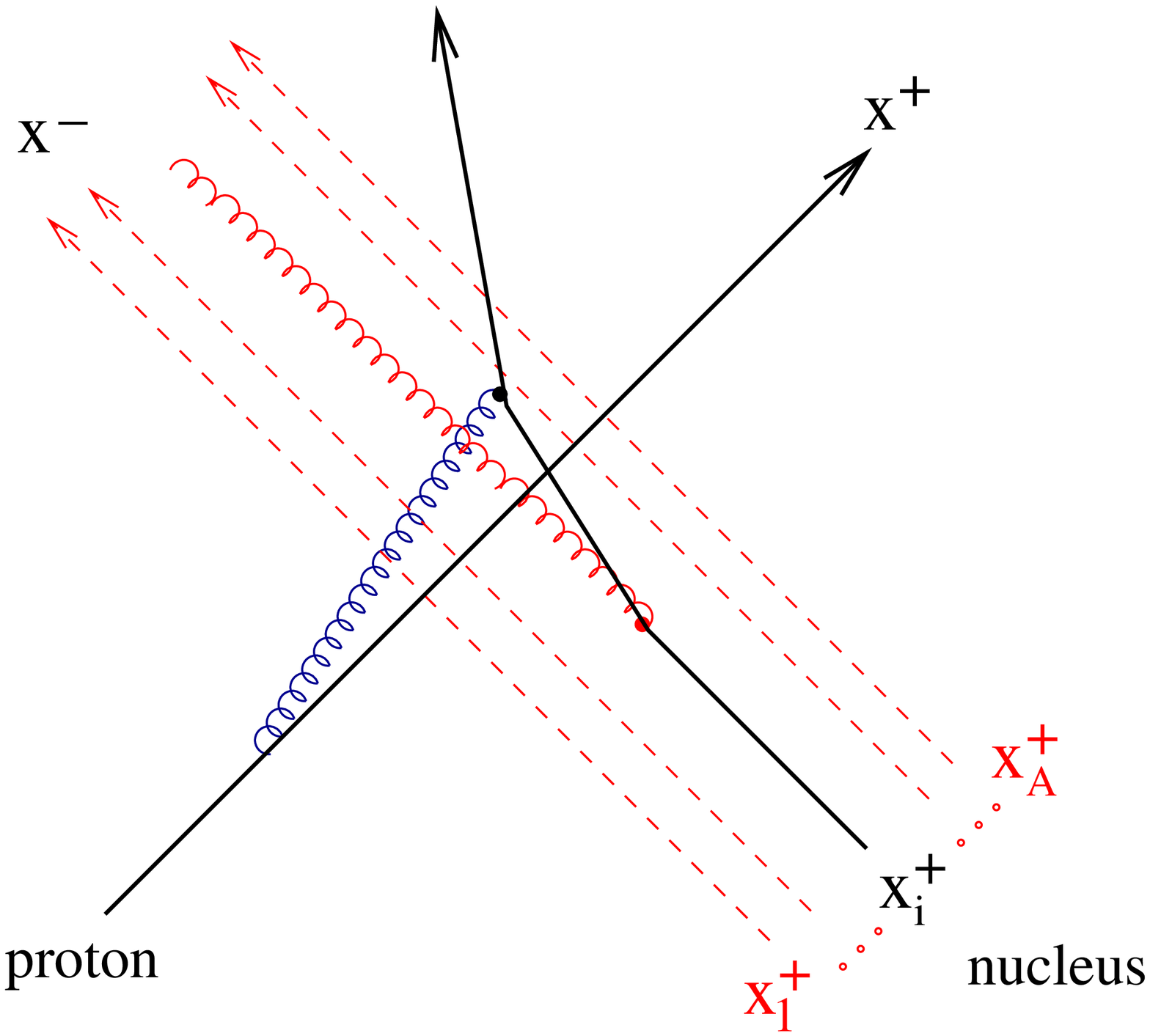}
\includegraphics[height=4cm,width=7cm]{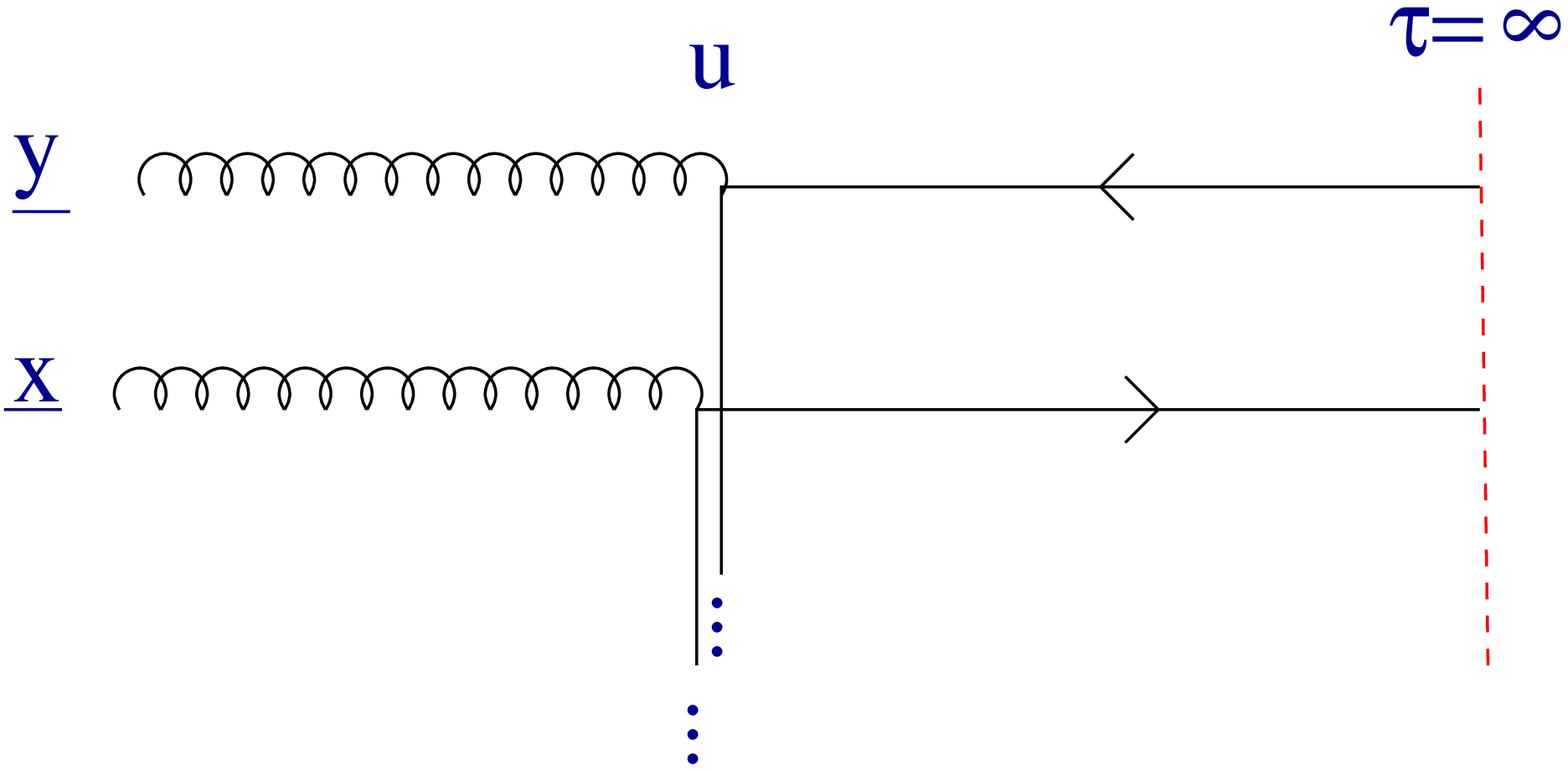}
\end{center}
\caption{Space-time picture of the diagram in \fig{f2} (left), and its mirror
reflection (right):}
\label{lc}
\end{figure}

The physics is clear: the proton has to emit a soft gluon before its
interaction with the nucleus. This emission is given by the soft-gluon
wavefunction of a valence quark
\beq
\psi^a_{\sigma,\lambda}(\ux,\uz) \, 
=\, 2\, g\, t^a \, \frac{i}{2\pi}
\frac{(\ux-\uz)\cdot\ud{\epsilon}^{\lambda}}{(\ux-\uz)^2}\, .
\label{gwf}
\eeq
The emitted soft gluon undergoes multiple rescatterings until being
absorbed by one of the valence quarks of the {\it i}th nucleon at the
longitudinal coordinate $u$. This way, the valence quark is kicked out
from its original light-cone trajectory towards the central rapidity
region and is likely to rescatter on the remaining nucleons in the
nucleus. A space-time picture of the collision is depicted in
\fig{lc}A. Integrating over the longitudinal coordinate $u$ and
summing over all nucleons in the nucleus one gets the following
factor:
\beq
\rho\int_0^L \!\!du\, e^{-(\ux-\uy)^2Q_{sg}^2\ln(1/|\ux-\uy| \Lambda)u/4L}\,\hat{V}
(\ux,\uy)\,e^{-(\ux-\uy)^2Q_{sq}^2\ln(1/|\ux-\uy| \Lambda)(L-u)/4L}.
\label{ei4}
\eeq
The first exponential term in \eq{ei4} accounts for the
propagation of a gluon dipole until the  longitudinal coordinate $u$ (see
\fig{lc}B) with 
\beq
Q_{sg}^2=4\pi\alpha_s^2\rho T_A(\ub)
\label{gss}
\eeq

\noi the gluon dipole saturation scale. The factor $\hat{V}(\ux,\uy)$ is 
the Fourier transform of the ${Gq\rightarrow qG}$ cross-section in the
high energy limit (see Appendix B for a detailed derivation):
\beq
\hat{V}(\ux,\uy)=N_c \, \int\! d^2 {l}\,e^{i\ud{l}\cdot(\ux-\uy)}
\frac{d\sigma^{Gq\rightarrow qG}}{ d^2 {l}}=\pi\alpha_s^2C_F \frac{1}{\hat{s}}
\ln\frac{1}{(\ux-\uy)^2\Lambda^2},
\label{v}
\eeq
where $\hat{s}=2k^+p_2^- = |{\un k}| \sqrt{s} \, e^{y - \frac{Y}{2}}$. 
The factor of $N_c$ entering the definition of $\hat{V}$ accounts for 
the three valence quarks in a nucleon. Finally, the second exponential
term in \eq{ei4} corresponds to the propagation of a quark dipole from
$u$ till the edge of the nucleus $L$, which is equal to $T_A(b)$ evaluated at
a fixed value of the impact parameter. Thus, the contribution to the
cross section of the diagram in \fig{f2} is:
\ben
\frac{d\sigma^b}{d^2 k\,dy} \, = \, \frac{1}{(2\pi)^2} \, \int\! d^2 x\,d^2 y\,d^2 z\, 
e^{i\uk(\ux-\uy)} \, \frac{dn^g}{d^2 z \,dy}
\een
\beq
\times \, \left[\rho \int_0^L \!\!du\, 
e^{-(\ux-\uy)^2Q_{sg}^2\ln(1/|\ux-\uy| \Lambda)u/4L}\,\hat{V}
(\ux,\uy)\,e^{-(\ux-\uy)^2Q_{sq}^2\ln(1/|\ux-\uy| \Lambda)(L-u)/4L}\right]\,,
\label{dsb}
\eeq
\noi where 
\beq
\frac{dn^g}{d^2 z\,dy}=\frac{\bar{\alpha}_s}{\pi} \, \frac{(\ux-\uz)\cdot(\uy-\uz)
}{(\ux-\uz)^2(\uy-\uz)^2},
\label{qdf2}
\eeq
\noi is the soft gluon distribution of the valence quark.

Rewriting $\alpha$ in \eq{dsa} and $\beta$ in \eq{dsb} using
Eqs. (\ref{alpha}), performing the integral over the longitudinal
coordinate $u$ in \eq{dsb} and shifting the integration variables in
the following way
\bea
\ux-\uz\equiv \ux,\\
\uy-\uz\equiv \uy,\\
\uz\equiv \ub,
\eea
the final result for the soft valence quark production cross section
in the quasi-classical approximation can be rewritten as:
\beq
\frac{d\sigma}{d^2 k\,dy} \, = \, \frac{d\sigma^a}{d^2 k\,dy}+
\frac{d\sigma^b}{d^2 k\,dy},
\label{dsf}
\eeq 
with
\begin{gather}
\frac{d\sigma^a}{d^2k\,dy}=\frac{1}{(2\pi)^2}
\frac{\bar{\alpha}_s \, N_c}{ 2\pi}e^{y-Y/2} \, \frac{|\uk|}{\sqrt{s}}
\int\! d^2 x\,d^2 y\,d^2 b\, 
e^{i\uk(\ux-\uy)}\,\frac{\ux\cdot\uy}{\ux^2\uy^2}\nonumber\\
\times \, \left[e^{-(\ux-\uy)^2Q_{sq}^2\ln(1/|\ux-\uy|\Lambda)/4}-
e^{-\ux^2Q_{sq}^2\ln(1/|\ux|\Lambda)/4}-
e^{-\uy^2Q_{sq}^2\ln(1/|\uy|\Lambda)/4}+1\right],
\label{dsaf}
\end{gather}
and

\begin{gather}
\frac{d\sigma^b}{d^2 k\,dy}= \frac{1}{(2\pi)^2} 
\frac{2\bar{\alpha}_s}{\pi} \frac{C_F}{1-\frac{C_F}{ N_c}}
\frac{e^{-y+Y/2}}{\sqrt{s}|\uk|} \, \int\! d^2 x\,d^2 y\,d^2 b\,
e^{i\uk(\ux-\uy)}\,\frac{\ux\cdot\uy}{\ux^2\uy^2} \, \frac{1}{(\ux-\uy)^2}
\nonumber\\
\times \, \left[e^{-(\ux-\uy)^2Q_{sq}^2\ln(1/|\ux-\uy|\Lambda)/4}
-e^{-(\ux-\uy)^2Q_{sg}^2\ln(1/|\ux-\uy|\Lambda)/4}\right].
\label{dsbf}
\end{gather}
In arriving at \eq{dsbf} we have used the definition of the saturation
scale from Eqs. (\ref{qss}) and (\ref{gss}) with $T_A(b)$ replaced by
$L$.
 
Equations (\ref{dsf}), (\ref{dsaf}) and (\ref{dsbf}) provide the
result for the small-x valence quark production in pA collisions at
high energies in the quasi-classical approximation.  


\subsection{Properties of the Quasi-Classical Cross Section}


In this Section we explore some of the properties of the valence quark
production cross section derived in the previous Section.

First let us consider the case of $k_T \ll Q_s$, where $k_T = |\un
k|$. For momenta of the produced quark much smaller than the
saturation scale, $k_T \ll Q_s$, the dynamics is governed by the
strong nuclear effects. In this region the integrals in Eqs.
(\ref{dsaf}) and (\ref{dsbf}) are dominated by large distances and
setting the logarithms in the exponentials to a constant value (more
precisely to $1$) is a good approximation. This way the integrals are
analytically doable, yielding
\beq
\frac{d\sigma^a}{d^2 k\,dy}=\frac{\bar{\alpha_s}N_c}{
  2\pi}\frac{e^{y-Y/2}}{\sqrt{s}}\int\! d^2 {b}\,
\left\{-\frac{1}{|\uk|}+\frac{2}{|\uk|}e^{-\uk^2/Q_{sq}^2}+
|\uk| \frac{e^{-\uk^2/Q_{sq}^2}}{Q_{sq}^2}\left[\mbox{Ei}\left(
\frac{\uk^2}{Q_{sq}^2}\right)+\ln \frac{Q_{sq}^4}{4\uk^2 \Lambda^2}\right]\right\},
\label{ska}
\eeq
\noi and
\begin{gather}
\frac{d\sigma^b}{d^2 k\,dy}= \frac{\bar{\alpha_s}}{ 2\pi}
\frac{C_F}{1-\frac{C_F}{N_c}} \, \frac{e^{-y+Y/2}}{\sqrt{s}}
\frac{1}{|\uk|}\int\! d^2\ud{b}\left\{
\left(\ln\frac{Q_{sg}^2}{4\Lambda^2}+
\gamma_E\right)\left[\Gamma\left(0,\frac{\uk^2}{Q_{sg}^2}\right)+
\ln\frac{\uk^2}{Q_{sg}^2}\right]\right.\nonumber \\
-\left.\left(\ln\frac{Q_{sq}^2}{4\Lambda^2}+
\gamma_E\right)\left[\Gamma\left(0,\frac{\uk^2}{Q_{sq}^2}\right)+
\ln\frac{\uk^2}{Q_{sq}^2}\right]\right\},
\label{skb}
\end{gather}
where Ei is the exponential integral function and $\Gamma$ is the
incomplete gamma function.

On the other hand, for very large transverse momenta, $k_T \gg Q_s$,
the nuclear effects are absent (tend to disappear) and the
perturbative result should be recovered. Expanding the exponentials in
\eq{dsa} and \eq{dsb} one gets the following expansion in powers of 
$Q_{sq}/k_T$ (``twists''):
\begin{gather}
\frac{d\sigma^a}{d^2 k\,dy}=\frac{\bar{\alpha_s}N_c }{2\pi}
\frac{e^{y-Y/2}}{\sqrt{s}} \int\! d^2 {b} \frac{Q_{sq}^2}{k_T^3}\left\{ 
\left[\ln\frac{\uk^2}{4\Lambda^2}+2\gamma_E-1\right]+ \frac{Q_{sq}^2}{k_T^2}
\left[\frac{3}{2}\ln^2\frac{\uk^2}{4\Lambda^2}\right.\right.\nonumber\\
\left.\left.+2\ln\frac{\uk^2}{4\Lambda^2}
(3\gamma_E-4)+6\gamma_E^2-16\gamma_E+\frac{29}{4}\right]+\dots\right\},
\label{lka2}
\end{gather}
\noi and
\begin{gather}
\frac{d\sigma^b}{d^2 k\,dy}=\frac{\bar{\alpha_s}N_c}{2\pi} \, \frac{e^{-y+Y/2}
}{\sqrt{s}}\int d^2 {b}\, \frac{Q_{sq}^2}{k_T^3}\left\{
\left[\ln\frac{\uk^2}{4\Lambda^2}+2\gamma_E\right]+\frac{Q_{sq}^2}{k_T^2}
\left(\frac{N_c}{C_F}+1\right)
\left[\frac{3}{8}\ln^2\frac{\uk^2}{4\Lambda^2}
\right.\right.\nonumber\\
\left.\left.
+\frac{3}{2}\ln\frac{\uk^2}{4\Lambda^2}(\gamma_E-1)+\frac{2}{3}
(2\gamma_E^2-4\gamma_E+1)\right]+\dots\right\}.
\label{lkb2}
\end{gather}
Remarkably, at large transverse momentum and fixed rapidity the
valence quark production cross section scales like $\sim 1/k_T^3$,
which is to be compared with the $\sim 1/k_T^4$ scaling of the
inclusive gluon production cross section. As is well-known, the quark
production cross section turns out to be more sensitive to the
ultraviolet.

\begin{figure}[ht]
\begin{center}
\includegraphics[height=8cm,width=11.5cm]{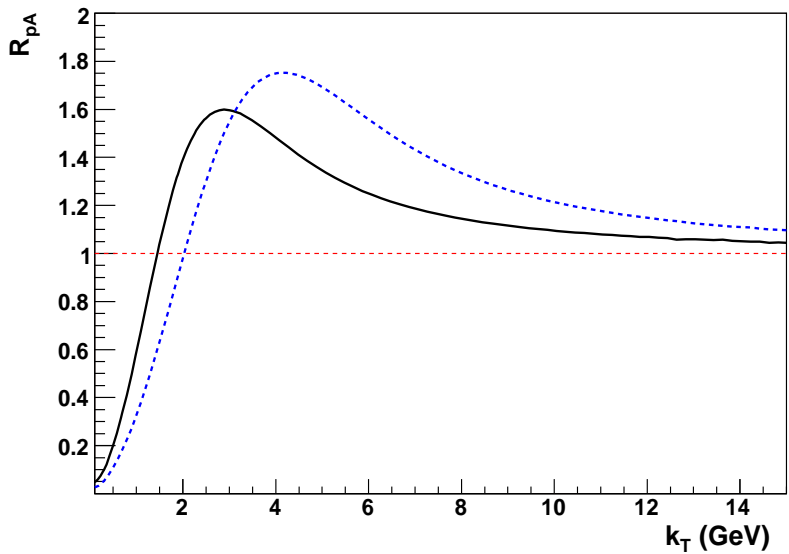}
\end{center}
\caption{Nuclear modification factor at central rapidity for $Q_{sq}^2=1$~GeV$^2$ 
(solid) and $Q_{sq}^2=2$~GeV$^2$ (dashed). The cutoff is
$\Lambda=0.1\, Q_{sq}$.}
\label{cronin}
\end{figure}

To study the transition between the small and large transverse
momentum regions and to have a quantitative measure of the nuclear
effects we introduce the nuclear modification factor, equal to the
ratio of the valence quark production cross section in $pA$ collisions
over the one in $pp$ collisions scaled by the number of binary
collisions
\beq
R_{pA}(y,k_T)=\frac{\frac{d\sigma^{pA}}{d^2 k\,dy}}{ 
A \, \frac{d\sigma^{pp}}{d^2 k\,dy}}\,.
\label{rpa}
\eeq
The cross section for proton-proton collisions can be obtained as a
particular case of our calculation, just keeping the first terms in
the twist expansions of Eqs. (\ref{lka2}) and (\ref{lkb2}) and setting
$A=1$. We get
\beq
A\, \frac{d\sigma^{pp}}{d^2 k\,dy} \, \approx \,
\frac{\bar{\alpha_s}N_c}{\pi} \, 
\frac{\cosh(y-Y/2)}{\sqrt{s}}\int\! d^2 {b} \, \frac{Q_{sq}^2}{k_T^3} \,
\ln\frac{\uk^2}{4\Lambda^2}\,,
\label{spp}
\eeq
where we have used $Q_{sq}^2\sim A^{1/3}$ and the fact that, for a
cylindrical nucleus, the integral over impact parameter gives just a
factor of the transverse area of the nucleus, $S_A\sim A^{2/3}$. In
what follows we will restrict our discussion about $R_{pA}$ to the
central rapidity region, $y=Y/2$. The limiting behavior of $R_{pA}$ at
momenta $k_T\ll Q_{sq}$ can be obtained by expanding \eq{ska} and
\eq{skb} in powers of $k_T/Q_{sq}$, yielding
\beq
R_{pA}(y=Y/2,k_T\ll Q_{sq})\approx 
\frac{\uk^2}{Q_{sq}^2}+\frac{\uk^4}{Q_{sq}^4}
\bigg( \gamma_E+\ln\frac{Q_{sq}^2}{4\, \Lambda^2}+
\frac{N_c}{1-\frac{C_F}{N_c}}\, \ln\frac{N_c}{C_F}-2 \bigg)+\dots \lt 1,
\label{rpask}
\eeq
indicating that deep in the saturation region the valence 
quark production is suppressed due to nuclear effects and that $R_{pA}$ 
increases with increasing momentum. On the other hand, from 
\eq{lka2}, \eq{lkb2} and \eq{spp} we derive the limiting behavior at large
momentum, $k_T\gg Q_{sq}$,
\beq
R_{pA}(y=Y/2,k_T\gg Q_{sq})\approx 1+
\frac{15}{8}\frac{Q_{sq}^2}{\uk^2}\ln^2\frac{\uk^2}{4\Lambda^2}
+\dots \gt 1,
\label{rpalk}
\eeq
which shows that $R_{pA}$ approaches unity from above at
asymptotically large transverse momentum. Therefore $R_{pA}$ must
reach a maximum value at intermediate momentum $k_T\sim Q_{sq}$, as it
is shown in \fig{cronin}, where we plot the nuclear modification
factor $R_{pA}$ at mid-rapidity ($y=Y/2$) built up by evaluating the
integrals in \eq{dsaf} and \eq{dsbf} numerically for two different
values of the quark saturation scale.

We therefore conclude that, in the framework of the quasi-classical
approximation, the nuclear modification factor for small-$x$ valence
quark production in the central rapidity region is less than 1 at
$k_T\lt Q_{sq}$ and has a Cronin enhancement at high $k_T\sim
Q_{sq}$. An equivalent (analogous) conclusion could be drawn for
values of rapidity outside the central region. However, at more
forward rapidities the nuclear wavefunction is probed at smaller
values of Bjorken-$x$, where the effects of quantum corrections become
important and may significantly alter this conclusion. This is known
to be the case for inclusive particle production in pA collisions, for
which the Cronin enhancement observed at central rapidity is
completely washed out by quantum evolution, turning it into a relative
suppression at forward rapidities, as was predicted for gluon
production in \cite{KKT,AAKSW} and experimentally confirmed in
\cite{Arsene:2003yk}.


\section{Including Quantum Evolution}

In this Section we are going to introduce the small-$x$ evolution
corrections into Eqs. (\ref{dsaf}) and (\ref{dsbf}). Since we are
interested in valence quark production, the small-$x$ evolution
corrections to Eqs. (\ref{dsaf}) and (\ref{dsbf}) will include both
the contributions from non-linear evolution equations for gluon
evolution \cite{BFKL,dip,yuri,bal,JKLW,FILM} and for reggeon evolution
\cite{KL,Kirschner1,Kirschner2,GR,IKMT} (with quarks in $t$-channel).


\subsection{Valence Quarks from the Proton}

We begin by considering case (a) above, in which the valence quark
originates in the incoming proton. For the quasi-classical case the
diagrams of this process are shown in \fig{f1}, with the expression
for the production cross section given by \eq{dsaf}. We begin by
rewriting \eq{dsaf} with the help of \eq{ei1} as
\begin{gather}
\frac{d\sigma^a}{d^2k\,dy}=\frac{1}{(2\pi)^2}
\frac{\bar{\alpha}_s \, N_c}{ 2\pi}e^{y-Y/2} \, \frac{|\uk|}{\sqrt{s}}
\int\! d^2 x\,d^2 y\,d^2 b\, 
e^{i\uk(\ux-\uy)}\,\frac{(\ux - \un b )\cdot (\uy - \un b)}{|\ux - \un
b|^2 \, |\uy - \un b|^2} \nonumber\\
\times \, \left[ N(\un x, \un b) + N (\un y, \un b) - N (\un x, \un y) \right].
\label{dsaf2}
\end{gather}

Inclusion of quantum evolution corrections will proceed along the
lines of \cite{KT} (see also \cite{JMK,JMK1,Marquet,yuri05,KT2}). The
evolution is divided into two components: emissions of gluons harder
than the produced valence quark (with rapidities greater than $y$) and
the emission of gluons softer than the valence quark (with rapidities
smaller than $y$). The emission of harder gluons leads to diagrams
with quarks in the $t$-channel, leading to reggeon evolution equations
resumming double leading logarithms of energy (powers of $\as \ln^2
s$) \cite{KL,Kirschner1,Kirschner2,GR,IKMT}. Here we will consider the
reggeon evolution in the transverse coordinate-space formalism
developed in \cite{IKMT}. The building block of that evolution is a
decay of a valence quark into a hard gluon and a much softer valence
quark.

Similar to \cite{KT} one can show that the incoming valence quark in
the proton may evolve by emitting hard gluons only before the
interaction with the target (at light cone times $\tau <0$). Only one
such emission in the final state after the interaction (at $\tau >0$)
is allowed: more final state emission would not generate longitudinal
logarithms needed for the double-logarithmic approximation to the
reggeon evolution employed here and in \cite{IKMT}. Thus the effect of
harder gluon emissions is to ``prepare'' a valence quark which then
will serve as an incoming quark in \fig{f1}, and, in turn, may split
into a soft quark and a hard gluon either before or after the
interaction, as shown in \fig{f1}.

To include the effects of harder gluon emissions we first have to
consider a slightly more sophisticated model of a proton. Up until now
we have modeled the projectile proton by an incoming valence quark. To
be able to construct a transverse coordinate analogy of the reggeon
evolution equation, as was done in \cite{IKMT}, we will model the
incoming proton by a dipole consisting of a quark at a transverse
space position $\un z_1$ and an anti-quark at $\un z_0$. The dipole
size $\un z_{10} \equiv \un z_1 - \un z_0$ is comparable to the
proton's transverse size $\sim 1/\Lambda_{QCD}$. If the quark in the
dipole carries a fraction $\alpha_1$ of its longitudinal (plus)
momentum component, the total rapidity interval in the dipole-nucleus
scattering is defined by $Y = \ln (s \, z_{10}^2)$ and the rapidity of
the valence quark in the wave function is $y_1 = \ln (\alpha_1 \, s \,
z_{10}^2)$ \cite{IKMT}, where $s$ is the center of mass energy of the
proton-nucleus system. \eq{dsaf2} now becomes
\begin{gather}
\frac{d\sigma^a}{d^2k\,dy} (\un z_{10}, \alpha_1) \, = \, \frac{1}{(2\pi)^2}
\frac{\bar{\alpha}_s \, N_c}{ 2\pi} \, \frac{\alpha}{\alpha_1}
\int\! d^2 x\,d^2 y\,d^2 z_1\, 
e^{i\uk(\ux-\uy)}\,\frac{(\ux - \un z_1 )\cdot (\uy - \un z_1)}{|\ux - \un
z_1|^2 \, |\uy - \un z_1|^2} \nonumber\\
\times \, \left[ N(\un x, \un z_1) + N (\un y, \un z_1) - N (\un x, \un y) \right],
\label{dsaf3}
\end{gather}
where $\alpha$ is defined by \eq{alpha}. Indeed to obtain the valence
quark production cross section one has to convolute the cross section
in \eq{dsaf3} with the dipole's light cone wave function squared
integrating over $\un z_{10}$ and $\alpha_1$. Such wave function is
well-known for the case of deep inelastic scattering (DIS) on the
nucleus \cite{NZ}. Thus, rigorously speaking, our discussion in this
Section will be most relevant for DIS. However, one could use the
formula we obtain below for proton-nucleus scattering by putting
$z_{10} \approx 1/\Lambda_{QCD}$ in $Y$ and by using $\alpha_1 \approx
1/3$.

\begin{figure}[ht]
\begin{center}
\includegraphics[width=15cm]{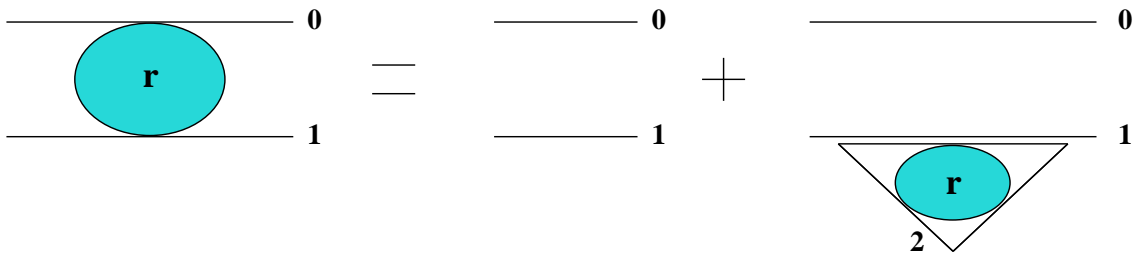}
\end{center}
\caption{Graphical representation of the evolution equation (\ref{rev}) in the text. }
\label{revfig}
\end{figure}

We denote by $r ({\un z}_{0}, {\un z}_1, \alpha_1; \un z,
\alpha)/\alpha_1$ the probability density of finding a soft valence quark at
transverse coordinate $\un z$ carrying longitudinal momentum fraction
greater than or equal to $\alpha$ in the wave function of a dipole
$\un z_{10}$ with the initial longitudinal momentum fraction carried
by the quark being $\alpha_1$.  The quantity $r$ obeys the following
evolution equation in the large-$N_c$ limit
\beq\label{rev}
r ({\un z}_{0}, {\un z}_1, \alpha_1; \un z, \alpha) \, = \, \delta^2
(\un z - \un z_1) \, + \,
\frac{\bas}{2 \, \pi} \, \int_{\alpha}^{\alpha_1 \,
\mbox{min} \{1, z_{01}^2/z_{21}^2 \}} \, \frac{d \alpha'}{\alpha'} \, \frac{d^2
z_2}{z_{12}^2} \, r ({\un z}_1, {\un z}_2, \alpha'; \un z, \alpha). 
\eeq
\eq{rev} is illustrated in \fig{revfig}. 
One can immediately see that \eq{rev} is a linearized version of the
nonlinear Eq. (43) in \cite{IKMT}. In \fig{revfig} the incoming quark
$1$ in the dipole splits into a hard gluon $1$ (denoted by a double
line on the right hand side of \fig{revfig}) and into a soft quark
$2$. The subsequent evolution continues in the new dipole $12$. In the
non-linear evolution equation derived in \cite{IKMT} the evolution
could also continue in the original dipole $01$ in the sense of gluon
(BK) evolution \cite{yuri,bal}. However, for an inclusive quantity,
such as valence quark production considered here, all further
interactions of dipole $01$ cancel by real-virtual cancellations. This
is similar to the case of inclusive gluon production cross section
considered in \cite{KT}.

The upper limit of the integral over $\alpha'$ (the longitudinal
momentum fraction of the soft valence quark at $\un z_2$) in \eq{rev}
results from two considerations: on the one hand, according to the
rules of light-cone perturbation theory \cite{BL} $\alpha' \le
\alpha_1$, while, on the other hand, rapidity ordering of the produced 
dipoles requires \cite{IKMT} 
\beq
\alpha' \, \ll \, \alpha_1 \, \frac{z_{10}^2}{z_{12}^2}
\eeq
as the rapidity of the soft valence quark in dipole $12$ is $y' = \ln
(\alpha' s \, z_{12}^2)$. (To understand this last formula for $y'$
remember that, in transverse momentum space, $y - (Y/2) = \ln (\alpha
\, \sqrt{s} / k_T)$ and $Y = \ln (s / k_T^2)$ leading to $y = \ln
(\alpha \, s / k_T^2)$.)

For the dipole-nucleus scattering the effect of harder gluons can be
included in the production cross section (\ref{dsaf3}) by
\begin{gather}
\frac{d\sigma^a}{d^2k\,dy} (\un z_{10}, \alpha_1) \, = \, \frac{1}{(2\pi)^2}
\frac{\bar{\alpha}_s \, N_c}{ 2\pi} \, \frac{\alpha}{\alpha_1} \,
\int\! d^2 x\,d^2 y\,d^2 z_1 \, d^2 z \,
e^{i\uk(\ux-\uy)}\, r ({\un z}_{0}, {\un z}_1, \alpha_1; \un z, \alpha) 
\, \frac{(\ux - \un z )\cdot (\uy - \un z)}{|\ux - \un
z|^2 \, |\uy - \un z|^2} \nonumber\\
\times \, \left[ N(\un x, \un z) + N (\un y, \un z) - N (\un x, \un y) \right],
\label{dsaf4}
\end{gather}
where we use $y = \ln ( \alpha \, s / k_T^2)$ to relate $\alpha$ and
$y$.

Inclusion of the evolution effects due to emissions of softer gluon
having rapidity less than $y$ is quite straightforward and goes
exactly parallel to \cite{KT,JMK,JMK1,Marquet,yuri05,KT2}. To include
these softer emissions we need to replace
\beq
N (\un x, \un y) \, \rightarrow \, N (\un x, \un y, y)
\eeq
in \eq{dsaf4}, where now $N (\un x, \un y, y)$ has to be found from
the non-linear evolution equation \cite{yuri,bal}

\ben
\frac{\partial N ({\un x}_{0}, {\un x}_1, Y)}{\partial Y} \, = \, 
\frac{\as \, N_c}{2 \, \pi^2} \, 
\int d^2 x_2 \, \frac{x_{01}^2}{x_{20}^2 \, x_{21}^2} \, 
\left[ N ({\un x}_{0}, {\un x}_2, Y) + 
N ({\un x}_{2}, {\un x}_{1}, Y) - N ({\un
x}_{0}, {\un x}_1, Y) \right. 
\een
\beq\label{eqN}
- \left. N ({\un x}_{0}, {\un x}_{2}, Y) \, N ({\un x}_{2}, {\un x}_{1}, Y)
\right]
\eeq
with the initial condition
\beq
 N ({\un x}_{0}, {\un x}_1, Y=0) \, = \, N ({\un x}_{0}, {\un x}_1).
\eeq

The final answer for the valence quark production cross section
including quantum evolution with the valence quark coming from the
projectile proton reads
\begin{gather}
\frac{d\sigma^a}{d^2k\,dy} (\un z_{10}, \alpha_1) \, = \, \frac{1}{(2\pi)^2}
\frac{\bar{\alpha}_s \, N_c}{ 2\pi} \, \frac{\alpha}{\alpha_1} \,
\int\! d^2 x\,d^2 y\,d^2 z_1 \, d^2 z \, 
e^{i\uk(\ux-\uy)}\, r ({\un z}_{0}, {\un z}_1, \alpha_1; \un z, \alpha) 
\, \frac{(\ux - \un z )\cdot (\uy - \un z)}{|\ux - \un
z|^2 \, |\uy - \un z|^2} \nonumber\\
\times \, \left[ N(\un x, \un z, y) + N (\un y, \un z, y) - N (\un x, \un y, y) \right].
\label{dsaf5}
\end{gather}
Once again we remind the reader that here $Y \approx y_1 = \ln
(\alpha_1 s \, z_{01}^2)$ and $y = \ln ( \alpha \, s / k_T^2)$. When
integrating over $\un z$ in \eq{dsaf5} one has to keep in mind that
rapidity ordering requires that the rapidity of the quark at
transverse coordinate $\un z$, given by ${\tilde y} = \ln [\alpha \, s
\, (\un z - \un z_0)^2]$, is limited to the interval $y < {\tilde y} <
Y$, which imposes a constraint on the range of $\un z$-integration.

The essential ingredients of \eq{dsaf5} are the Reggeon exchange
amplitude $r$ obeying the evolution equation (\ref{rev}) which sums up
leading double logarithms $\as \, \ln^2 s$, and the amplitude $N$
obeying \eq{eqN} summing up usual leading logarithms $\as \, \ln s$.
We remind the reader that the fact that we are using the amplitude $r$
obeying a {\sl linear} evolution equation (\ref{rev}) to describe the
evolution between the produced valence quark and the projectile is not
an approximation but an exact result of {\sl cancellations} of
non-linear Reggeon evolution corrections from \cite{IKMT} in that
rapidity region.
 
If one is interested only in rapidity dependence of the cross section,
integration of \eq{dsaf5} over $k_T$ yields 
\begin{gather}
\frac{d\sigma^a}{dy} (\un z_{10}, \alpha_1) \, = \, 
\frac{\bar{\alpha}_s \, N_c}{\pi} \, \frac{\alpha}{\alpha_1} \,
\int\! d^2 x\,d^2 z_1 \, d^2 z \, r ({\un z}_{0}, {\un z}_1, \alpha_1; \un z, \alpha) 
\, \frac{1}{|\ux - \un z|^2} \, N(\un x, \un z, y),
\label{dsaf6}
\end{gather}
where we have also made use of the fact that $N(\un x, \un x, y) = 0$.
In the leading double logarithmic approximation we find $\alpha$ from $y \approx
\ln (\alpha \, s /Q_s^2 (y))$ where $Q_s^2 (y)$ is the quark saturation scale of the 
nucleus at rapidity $y$, which we use as the typical transverse
momentum of the produced quarks. $Q_s^2 (y)$ is rapidity-dependent due
to the effects of nonlinear evolution equation (\ref{eqN}).


\subsection{Properties of the Cross Section with Evolution}

First of all, let us point out that \eq{rev} can be solved
analytically if one averages it over the directions of $\un z_{01}$
(or $\un z_{21}$). Defining
\beq
\un \rho \, \equiv \, \un z - \un z_0
\eeq
we can rewrite \eq{rev} as
\beq\label{rev1}
r ({\un z}_{01}, \alpha_1; \un \rho, \alpha) \, = \, \delta^2
(\un \rho - \un z_{10}) \, + \,
\frac{\bas}{2 \, \pi} \, \int_{\alpha}^{\alpha_1 \,
\mbox{min} \{1, z_{01}^2/z_{21}^2 \}} \, \frac{d \alpha'}{\alpha'} \, \frac{d^2
z_2}{z_{12}^2} \, r ({\un z}_{21}, \alpha'; \un r, \alpha),  
\eeq
which, after angular averaging, becomes
\beq\label{rev2}
r ({z}_{01}, \alpha_1; \rho, \alpha) \, = \, \frac{1}{2 \, \pi \, \rho} \, \delta
(\rho - z_{10}) \, + \,
\frac{\bas}{2 \, \pi} \, \int_{\alpha}^{\alpha_1 \,
\mbox{min} \{1, z_{01}^2/z_{21}^2 \}} \, \frac{d \alpha'}{\alpha'} \, \frac{d^2
z_2}{z_{12}^2} \, r ({z}_{21}, \alpha'; r, \alpha). 
\eeq

Similar to \cite{IKMT} we first perform a Mellin transform 
\beq
r ({z}, \alpha_1; \rho, \alpha) \, = \, \int \frac{d
\omega}{2 \pi i} \, \left( \frac{\alpha_1 \, z^2}{\alpha \, \rho^2}
 \right)^\omega \, r_\omega (z, \rho)
\eeq
corresponding to Laplace transformation in rapidity. Here the
$\omega$-integration runs parallel to the imaginary axis to the right
of the origin and, due to rapidity ordering, we use $\alpha_1 \, z^2 >
\alpha \, \rho^2$. \eq{rev2} becomes
\beq
\omega \, r_\omega (z_{01}, \rho) \, = \, \frac{1}{2 \, \pi \, \rho} \, \delta
(\rho - z_{01}) \, + \, \frac{\bas}{2} \, \int_0^\infty \, \frac{d
z_{21}^2}{z_{21}^2} \, \left( \mbox{min} \left\{ \frac{z_{21}^2}{z_{01}^2}, 1 
\right\} \right)^\omega \, r_\omega (z_{21}, \rho).
\eeq
We perform another Mellin transformation
\beq
r_\omega (z, \rho) \, = \, \int \, \frac{d \lambda}{2 \pi i} \, \left(
\frac{\rho^2}{z^2} \right)^\lambda \, r_{\omega \, \lambda}
\eeq
with the $\lambda$-integration contour being the same as for the
$\omega$-integration above and assuming that $z < r$ for all
$z_{ij}$'s. We obtain
\beq 
\omega \, r_{\omega \, \lambda} \, = \, \frac{1}{\pi \, \rho^2} \, + \, 
\frac{\bas}{2} \, \frac{\omega}{\lambda \, (\omega - \lambda)} \,  r_{\omega \, \lambda}
\eeq
which gives
\beq
r_{\omega \, \lambda} \, = \, \frac{1}{\pi \, \rho^2} \, \frac{\lambda \, 
(\omega - \lambda)}{\omega \, [\lambda \, \omega - \lambda^2 - \frac{\bas}{2}]}.
\eeq
The solution of \eq{rev2} is
\beq\label{rsol}
r ({z}_{01}, \alpha_1; \rho, \alpha) \, = \, \frac{1}{\pi \, \rho^2} \, \int \frac{d
\omega}{2 \pi i} \, \frac{d \lambda}{2 \pi i} \, 
\left( \frac{\alpha_1 \, z_{01}^2}{\alpha \, \rho^2} \right)^\omega \, \left(
\frac{\rho^2}{z_{01}^2} \right)^\lambda \, \frac{\lambda \, 
(\omega - \lambda)}{\omega \, [\lambda \, \omega - \lambda^2 -
\frac{\bas}{2}]}.
\eeq

Similar to \cite{IKMT} we can first perform the integration over
$\omega$ in \eq{rsol}: arguing that the high-energy asymptotics is
dominated by the rightmost pole in $\omega$ plane and picking that
pole yields
\beq\label{rsol1}
r ({z}_{01}, \alpha_1; \rho, \alpha) \, \approx \, \frac{1}{\pi \, \rho^2} \, \int 
\frac{d \lambda}{2 \pi i} \, 
\left( \frac{\alpha_1 \, z_{01}^2}{\alpha \, \rho^2} \right)^{\lambda + 
\frac{\bas}{2 \, \lambda}}
 \, \left(\frac{\rho^2}{z_{01}^2} \right)^\lambda \,
\frac{\bas}{2 \, \lambda^2 + \bas}.
\eeq
\eq{rsol1} can be evaluated in the saddle point approximation near the 
saddle point at $\lambda = \sqrt{\bas /2}$, which gives
\beq\label{rsol2}
r ({z}_{01}, \alpha_1; \rho, \alpha) \, \propto \, \frac{1}{\rho^2} \,
e^{\sqrt{2 \, \bas} \, (y_1 - y)} \, \left(\frac{\rho^2}{z_{01}^2} 
\right)^{\sqrt{\frac{\bas}{2}}}
\eeq
leading to the well-known intercept of $\sqrt{2 \, \bas}$
\cite{KL,Kirschner1,Kirschner2}. A more careful evaluation of \eq{rsol1} 
can be done by direct integration. First rewrite \eq{rsol1} as
\beq\label{rsol3}
r ({z}_{01}, \alpha_1; \rho, \alpha) \, \approx \, \frac{1}{\pi \, \rho^2} \, \int 
\frac{d \lambda}{2 \pi i} \, 
\left( \frac{\alpha_1}{\alpha} \right)^{\lambda}  
\, \sum_{n=0}^\infty \, \frac{1}{n!} \left[ \frac{\bas}{2 \, \lambda} \, 
\ln \left( \frac{\alpha_1 \, z_{01}^2}{\alpha \, \rho^2} \right) \right]^n
 \, \frac{\bas}{2 \, \lambda^2 + \bas}.
\eeq
The expression obtained in \eq{rsol3} has poles at $\lambda=0$ and at
$\lambda = \pm i \sqrt{\bas/2}$. From \eq{rsol1} it is clear that the
poles at $\lambda = \pm i \sqrt{\bas/2}$ do not give leading high
energy asymptotics. Concentrating on poles at $\lambda=0$ we neglect
$2 \, \lambda^2$ in the $2 \, \lambda^2 + \bas$ denominator of
\eq{rsol3}, integrate each or the terms in the series by picking up
the pole at $\lambda=0$ and resum back the series obtaining
\beq\label{rsol4}
r ({z}_{01}, \alpha_1; \rho, \alpha) \, \approx \, \frac{1}{\pi \, \rho^2} \,  
\sqrt{\frac{\bas \, \ln \left( \frac{\alpha_1 \, z_{01}^2}{\alpha 
\, \rho^2} \right)}{2 \, \ln \frac{\alpha_1}{\alpha} }} \, I_1 \left( 
\sqrt{2 \, \bas \, \ln \left( \frac{\alpha_1 \, z_{01}^2}{\alpha 
\, \rho^2} \right) \, \ln \frac{\alpha_1}{\alpha}} \right).
\eeq

Now we turn to studying the behavior of the nuclear modification
factor under quantum evolution. First we note that the cross section
in \eq{dsaf5} can be further simplified assuming transverse
translational invariance of the solutions of the non-linear evolution
equation for the dipole scattering amplitude, \eq{eqN}, which can be
achieved by considering an infinite and homogeneous nucleus in the
transverse plane. Under this assumption the dipole scattering
amplitude becomes a function of a single spatial variable, $N(\un
r,\ub,y)\rightarrow N(r,y)$, and, shifting the integration variables
in in the following way:
\bea
\ux-\uz\equiv \un v,\\
\uy-\uz\equiv \un w,\\
\uz_1\equiv \ub,
\eea
\eq{dsaf5} can be rewritten as
\begin{gather}
\frac{d\sigma^a}{d^2k\,dy} (\un z_{10}, \alpha_1) \, = \, \frac{1}{(2\pi)^2}
\frac{\bar{\alpha}_s \, N_c}{ 2\pi} \, \frac{\alpha}{\alpha_1} \,
\int\!d^2b\,d^2z\,r({\un z}_{0}, {\un z}_1, \alpha_1; \un z, \alpha) 
\int\! d^2 v\,d^2 w \, 
e^{i\uk(\un v-\un w)}\, 
\, \frac{\un v\cdot \un w }{|\un v|^2 \, |\un w|^2} \nonumber\\
\left[ N(|\un v|, y) + N (|\un w|, y) - N(|\un v-\un w|, y) \right].
\label{dsaf7}
\end{gather}

Noticeably, the integrations over the arguments of the quark
probability density $r$ obeying the linear evolution equation
(\ref{rev}) and the one over the arguments of the dipole scattering
amplitude $N$ decouple. Thus, the linear evolution (\ref{rev}) only
introduces to the cross section a rapidity-dependent prefactor, which
carries no information at all of the nucleus and is therefore
irrelevant for the nuclear modification factor. This prefactor can be
calculated using the approximate solution of the reggeon evolution
equation given in \eq{rsol4}. The limits of integration are set by the
condition $y_1 - y \ge \ln \frac{\alpha_1z_{01}^2}{\alpha z^2}=y_1 - {\tilde y} \ge 0$,
yielding
\beq
\int\!d^2z\,r ({\un z}_{0}, {\un z}_1, \alpha_1; \un z, \alpha) =
\frac{y_1 - y}{\ln\frac{\alpha_1}{\alpha}}I_2
\left(\sqrt{2\, \bas\, (y_1 - y) \, \ln\frac{\alpha_1}{\alpha}}\,\right). 
\eeq 

Finally, after performing analytically three of the integrals in the Fourier
transform, \eq{dsaf7} turns into 
\ben
\frac{d\sigma^a}{d^2k\,dy} (\un z_{10}, \alpha_1) \, = \, 
\int\,d^2b\, \frac{y_1 - y}{\ln\frac{\alpha_1}{\alpha}}I_2
\left(\sqrt{2\, \bas\, (y_1 - y) \, \ln\frac{\alpha_1}{\alpha}}\,\right) \,
\een
\beq
\times \, \int_0^{\infty}\!\! \!d v \,
\left[\frac{2}{k} \, J_1(k \, v) - v  \, J_0(k\, v) \, \ln\frac{1}{v\, \Lambda}\right] 
\, N(v, y). 
\label{dsaf8}
\eeq

\begin{figure}[ht]
\begin{center}
\includegraphics[width=12cm]{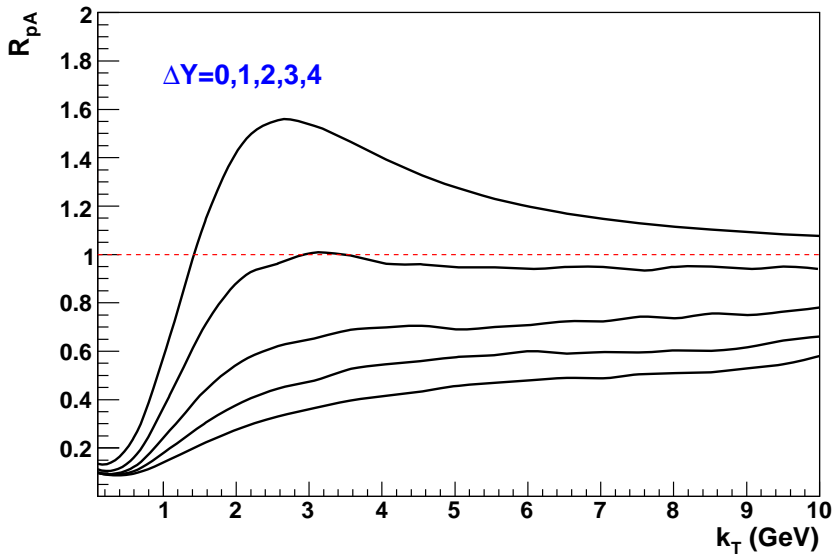}
\end{center}
\caption{Evolution of $R_{pA}(k_T,y)$ for valence quarks with rapidity. The initial condition
  corresponds to $Q_{sA}^2=1$~GeV$^2$ for a nucleus and
  $Q_{sp}^2=\Lambda^2=0.1$~GeV$^2$ for a proton. The rapidity
  interval is measured from mid-rapidity, $\Delta Y=y-Y/2$.}
\label{evcron}
\end{figure}

We now evaluate the nuclear modification factor $R_{pA}$ at increasing
values of rapidity towards the proton fragmentation region, where the
contribution to the cross section given in \eq{dsaf8} is the dominant
one. The dipole scattering amplitude entering this equation is
obtained by numerically solving \eq{eqN}, with initial conditions
given by the classical limit, \eq{ei1}, with $Q_{sA}^2\!=\!1$~GeV$^2$
for the nucleus and $Q_{sp}^2\!=\!0.1$~GeV$^2$ for the proton, and
putting $\alpha_s\!=\!0.2$.  The contribution to the cross section
corresponding to the valence quark coming from the nucleus is included
using the quasi-classical formula (\ref{dsbf}) at all values of
rapidity. This is a good approximation for forward enough rapidities,
where this contribution is vanishingly small anyway. The results for
$R_{pA}$ at rapidities $\Delta Y=y-Y/2=0,1,2,3,4$, are shown in
\fig{evcron}.

Our results show that the Cronin peak present in the initial condition
is wiped out by the evolution, turning into a relative suppression at
more forward rapidities. The rate of disappearance of the initial
enhancement is similar to the one found in \cite{AAKSW} for inclusive
gluon production. This is not an unexpected result, since the effects
of the reggeon evolution in the $pp$ and $pA$ cross sections are
divided out in the nuclear modification factor, whose rapidity
dependence is entirely driven by the dipole scattering amplitude, as
can be seen from \eq{dsaf8}. Moreover, the nuclear modification ratio
at large rapidity increases with increasing transverse momentum, and
eventually approaches unity at asymptotically large momentum.  The
slightly irregular behavior of the nuclear modification factor plotted
in \fig{evcron} at large momentum is due to the numerical uncertainty
associated to the Fourier transform in \eq{dsaf8} and has by no means
any physical origin.


\subsection{Valence Quarks from the Nucleus}

Now we want to include quantum evolution corrections into the
expression for valence quark production cross section given in
\eq{dsbf}, corresponding to the case (b) described above, in which the
valence quark originates in the nuclear wave function. As can be seen
from \fig{f2}, the inclusion of gluon emissions with rapidities
greater than the rapidity $y$ of the produced valence quark can be
straightforwardly accomplished using the linear dipole (BFKL)
evolution, similar to how it was done in \cite{KT} for gluon
production. 

We start by rewriting \eq{dsbf} for the case of dipole-nucleus
scattering, similar to how it was done above in Sect. 3.1 for the case
(a):
\begin{gather}
\frac{d\sigma^b}{d^2 k\,dy} (\un b_{01}) \, = \, \frac{1}{(2\pi)^2} 
\frac{2\bar{\alpha}_s}{\pi} \frac{C_F}{1-\frac{C_F}{ N_c}}
\frac{e^{-y+Y/2}}{\sqrt{s}|\uk|} \, \int\! d^2 x\,d^2 y\,d^2 B\,
e^{i\uk(\ux-\uy)}\,\sum_{i,j=0}^1 \, (-1)^{i+j} \, 
\frac{(\ux - \un b_i) \cdot (\uy - \un b_j)}{|\ux - \un b_i|^2 
\, |\uy - \un b_j|^2} 
\nonumber\\
\times \, \frac{1}{(\ux-\uy)^2} \, \left[e^{-(\ux-\uy)^2Q_{sq}^2\ln(1/|\ux-\uy|\Lambda)/4}
-e^{-(\ux-\uy)^2Q_{sg}^2\ln(1/|\ux-\uy|\Lambda)/4}\right].
\label{dsbf1}
\end{gather}
Here $\un b_0$ and $\un b_1$ are transverse coordinates of the quark
and the antiquark in the incoming dipole and $\un B = (1/2) (\un b_0 +
\un b_1)$ is the transverse coordinate of the center of the dipole. To 
include the quantum evolution corrections in the rapidity interval
between $y$ and $Y$ we first denote the number of dipoles with
transverse coordinates ${\un b}_{0'}, \un b_{1'}$ at rapidity $y$
generated by the evolution from the original dipole ${\un b}_{0}, \un
b_1$ having rapidity $Y$ by $n_1 ({\un b}_{0}, \un b_1; {\un b}_{0'},
\un b_{1'}; Y-y)$. This quantity is determined from the following evolution 
equation \cite{dip}
\ben
\frac{\partial n_1 ({\un b}_{0}, \un
b_1; {\un b}_{0'}, \un b_{1'}; y)}{\partial y} \, = \, 
\frac{\as \, N_c}{2 \, \pi^2} \, 
\int d^2 b_2 \, \frac{b_{01}^2}{b_{20}^2 \, b_{21}^2} \, 
\bigg[ n_1 ({\un b}_{0}, \un
b_2; {\un b}_{0'}, \un b_{1'}; y) + 
n_1 ( {\un b}_{2}, \un
b_1; {\un b}_{0'}, \un b_{1'}; y) 
\een
\beq\label{eqn}
- n_1 ({\un b}_{0}, \un b_1; {\un b}_{0'}, \un b_{1'}; y) \bigg] 
\eeq
with the initial condition 
\beq
n_1 ({\un b}_{0}, \un b_1; {\un b}_{0'}, \un b_{1'}; y=0) \, = \,
\delta ( \un b_0 - \un b_{0'} ) \, \delta ( \un b_1 - \un b_{1'} ).
\eeq
The inclusion of harder gluon emissions is accomplished by the
substitution \cite{KT}
\beq\label{sig_hard}
\frac{d\sigma^b}{d^2 k\,dy} (\un b_{01}) \, \rightarrow \, 
\int d^2 b_{0'} \, d^2 b_{1'} \, n_1 ({\un b}_{0}, \un b_1; 
{\un b}_{0'}, \un b_{1'}; Y-y) \, \frac{d\sigma^b}{d^2 k\,dy} 
(\un b_{0'1'}).
\eeq

\begin{figure}[ht]
\begin{center}
\includegraphics[width=12cm]{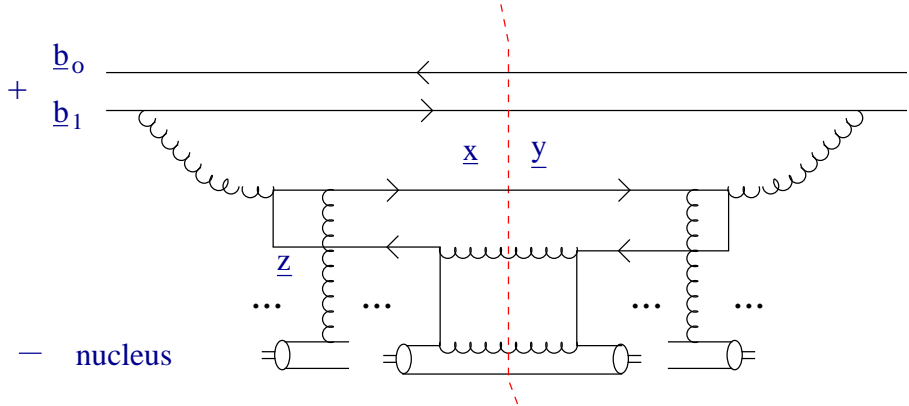}
\end{center}
\caption{Valence quark production in the case when the valence quark 
originates in the target nucleus with one rung of small-$x$ evolution
included. This diagram in turn serves as a lowest order diagram for
the inclusion of small-$x$ evolution corrections discussed in the
text.}
\label{evol1}
\end{figure}

Inclusion of softer gluon emissions in the rapidity interval between
$0$ and $y$ is somewhat more involved. The structure of the
quasi-classical result in \eq{dsbf} crucially depends on the fact that
the $t$-channel quark exchanges in \fig{f2} are
instantaneous. Attaching a single rung of Reggeon evolution to the
quark lines in \fig{f2} would immediately separate the quark lines
into a long-lived $s$-channel part and the short-lived $t$-channel
part, as shown in \fig{evol1}. Now the $s$-channel quark line allows
for inclusion of quantum evolution corrections as was done in \eq{rev}
and in \fig{revfig}. Thus it is the diagram in \fig{evol1} which gets
dressed by quantum corrections. The answer for valence quark
production cross section will consist of gluon production term given
by the diagram in \fig{f2} plus the diagram in
\fig{evol1}, which should be enhanced by quantum corrections. Indeed, the emissions 
of gluons with rapidities larger than the rapidity $y$ of the produced
quark are allowed in \fig{f2} and do not significantly modify its
structure. The real difference between Figs. \ref{f2} and
\ref{evol1} is due to evolution between the produced quark and the
target nucleus, which is allowed only in the latter. In fact, the
quasi-classical term from \eq{dsbf} and \fig{f2} will become
negligible compared to \fig{evol1} when quantum corrections are
included. The situation is similar to the interactions of a current
for dilaton-gluon coupling with the nucleus, as was first considered
in \cite{GM}. Since our expressions here should work for very high
energies, we will neglect the contribution of \fig{f2} compared to
that of \fig{evol1} after quantum evolution is included.

The calculation of the diagram in \fig{evol1} proceeds along the lines of
\cite{KT2}. The wave function squared for an incoming quark at transverse 
coordinate $\un 0$ decaying into a gluon which then decays into a hard
massless quark and a soft anti-quark is given by \cite{KT2}
\beq\label{Phi1}
{\tilde \Phi} (\un x, \un y ; \un z, \alpha) \, = \, 4 \, C_F \, 
\bigg(\frac{\as}{\pi}\bigg)^2 \, \frac{\un x \cdot (\un x - \un z)\ 
\un y \cdot (\un y - \un z) + \epsilon_{ij} \, x_i \, z_j \, \epsilon_{kl} \, 
y_k \, z_l}{\un x^2 \, \un y^2 \, (\un x - \un z)^2 \, (\un y - \un z)^2 }.
\eeq
The hard quark (to be produced) has transverse coordinates $\un x$ and
$\un y$ to the left and to the right of the cut correspondingly, as
shown in \fig{evol1}. A much softer anti-quark has a transverse
coordinate $\un z$ and a fraction $\alpha \ll 1$ of the gluon's
longitudinal momentum. Also, $\epsilon_{12} = 1$, $\epsilon_{21} = -1$
and $\epsilon_{11} = \epsilon_{22} =0$.

Since the emissions of harder gluons which are described above in the
dipole model framework lead to production of a dipole with transverse
coordinates which for the moment we will denote by $\un b_{0}$ and
$\un b_{1}$ for the quark and for the anti-quark, we rewrite the wave
function from \eq{Phi1} for the case of an incoming dipole (instead of
incoming quark) as
\beq\label{Phi2}
{\tilde \Phi} (\un x, \un y; \un z, \alpha) \, \rightarrow \,
\sum_{i,j=0}^1 \, (-1)^{i+j} \, 
\Phi (\un x - \un b_{i}, \un y - \un b_{j}; \un z, \alpha)
\eeq
where 
\ben
\Phi (\un x - \un b_{i}, \un y - \un b_{j}; \un z, \alpha) \, = \,  4 \, C_F \, 
\bigg(\frac{\as}{\pi}\bigg)^2 \, \frac{1}{(\un x - \un b_{i})^2 \, 
(\un y - \un b_{j})^2 \, (\un x - \un z)^2 \, (\un y - \un z)^2 }
\een
\beq
\times \, \left[ (\un x -\un b_{i} ) \cdot (\un x - \un z)\ 
(\un y - \un b_{j}) \cdot (\un y - \un z) + \epsilon_{mn} \, (\un x  - \un b_{i})_m \,
(\un x - \un z)_n \, \epsilon_{kl} \, (\un y  - \un b_{j})_k \, (\un y - \un z)_l \right].
\eeq

The interaction of the soft quark with transverse coordinate $\un z$
with the nucleus is limited to exchange of two $t$-channel quarks, as
shown in \fig{evol1}. Indeed, since this quark has the same transverse
coordinate on both sides of the cut, the multiple rescattering
corrections (with gluon exchanges like the ones shown in
Figs. \ref{f1} and \ref{f2}) cancel via real-virtual
cancellations. The interaction of the quark at $\un z$ with the
nucleus is then given by (see Appendix B of \cite{IKMT})
\beq\label{rintlo}
r_{int}^{(0)} (\un z, \alpha) \, = \, \frac{2 \, C_F^2 \, \pi \,
\as^2}{\alpha \, {\hat s}} \, \rho \, T_A (\un b) \, \ln \frac{\hat s}{\Lambda^2}
\eeq
where $\hat s = s \, e^{-Y+y}$ is the center-of-mass energy of the
incoming gluon-nucleus system in \fig{evol1} with $s$ the
center-of-mass energy of the whole collision. $\Lambda$ is some
infrared cutoff and we took into account the fact that nucleon
contains $N_c$ valence quarks.

The hard quark having transverse coordinates $\un x$ and $\un y$ in
\fig{evol1} also can interact with the nucleus. Since the transverse
coordinates of the hard quark are different on both sides of the cut
($\un x$ and $\un y$), real-virtual cancellations do not take place,
and the interaction of this quark with the target brings in a
Glauber-Mueller factor of
\beq\label{glamu}
e^{-(\ux-\uy)^2Q_{sq}^2\ln(1/|\ux-\uy|\Lambda)/4}.
\eeq

\begin{figure}[th]
\begin{center}
\includegraphics[width=10cm]{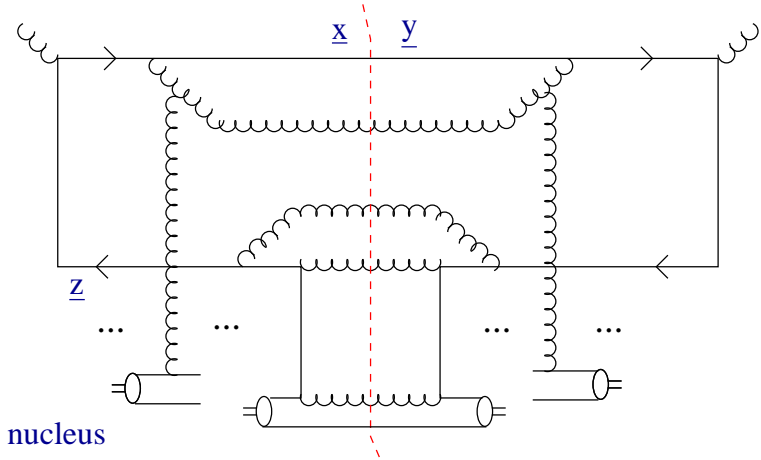}
\end{center}
\caption{Some examples of quantum evolution corrections to the 
diagram in \fig{evol1}: one may have emission and absorption by the
hard and soft quark lines, as shown in the figure and discussed in the
text.}
\label{evol2}
\end{figure}

Combining Eqs. (\ref{Phi2}), (\ref{rintlo}) and (\ref{glamu}) and
adjusting the overall normalization similar to how it was done in
\cite{KT2} we write the following expression for the quasi-classical 
valence quark production cross section in \fig{evol1}
\ben
\frac{d\sigma^b}{d^2 k\,dy} (\un b_{01}) \, = \, \frac{1}{2 \, (2 \pi)^4} \, 
\int d^2 x \, d^2 y \, d^2 z \, d^2 B \, e^{i \un k \cdot (\un x - \un y)} \,
\int_0^1 d \alpha \, 
\sum_{i,j=0}^1 \, \Phi (\un x - \un b_{i}, \un y - \un b_{j}; \un z, \alpha) 
\een
\beq\label{bev1}
\times \, 
r_{int}^{(0)} (\un z, \alpha) \,
e^{-(\ux-\uy)^2Q_{sq}^2\ln(1/|\ux-\uy|\Lambda)/4}.
\eeq

\eq{bev1} makes inclusion of quantum corrections into the diagram of \fig{evol1} 
quite straightforward. The emission of harder gluons is accomplished
by the substitution in \eq{sig_hard} as discussed above. All the
possible emissions of softer gluons are illustrated in \fig{evol2}. A
gluon may be emitted and absorbed by the harder quark, leading to
nonlinear small-$x$ evolution of the quark-anti-quark dipole $\un x$,
$\un y$. Similar to \cite{KT} such evolution is included in the
expression (\ref{bev1}) by substituting
\beq\label{brepl}
e^{-(\ux-\uy)^2Q_{sq}^2\ln(1/|\ux-\uy|\Lambda)/4} \, \rightarrow \, 1
- N (\un x, \un y, y)
\eeq
where $N (\un x, \un y, y)$ is given by the solution of \eq{eqN}. In
making the substitution of \eq{brepl} we are including gluon emissions
both before and after the interaction with the target \cite{KT}.

A gluon may {\sl not} be emitted by the hard quark at $\un x$ or $\un
y$ and absorbed by the soft anti-quark at $\un z$ (and vice versa) because
such diagrams are $N_c$-suppressed. The only other allowed $s$-channel
gluon lines are the emissions and absorptions of the gluon by the soft
anti-quark at $\un z$ as is also shown in \fig{evol2}. The gluon emitted
this way can only be emitted before the interaction with the target:
late-time emissions cancel by real-virtual cancellations
\cite{MC}. Since these gluons are emitted before the interaction on 
both sides of the cut, and as their transverse momenta are integrated
over, their interactions with the target nucleus will cancel, just
like they did for the anti-quark at $\un z$. The result of such gluon
emissions would thus be a linear evolution similar to the one shown in
\fig{revfig} but with $r_{int}^{(0)} (\un z, \alpha)$ from \eq{rintlo} 
as the initial condition. The evolution equation reads
\beq\label{rintev}
r_{int} (\un x, \un y ; \un z, \alpha) \, = \, r_{int}^{(0)} (\un z, \alpha) \, + \,
\frac{\bas}{2 \, \pi} \, \int_{\alpha_0}^{\alpha \,
\mbox{min} \left\{1, \frac{(\un z - \un x)^2}{(\un z - \un z_2)^2}, 
\frac{(\un z - \un y)^2}{(\un z - \un z_2)^2} \right\}} \, 
\frac{d \alpha'}{\alpha} \, \frac{d^2
z_2}{(\un z - \un z_{2})^2} \, r_{int} ({\un z}, \un z;  {\un z}_2, \alpha'). 
\eeq
Just like for \eq{rev} the upper limit of $\alpha'$-integral is
derived by requiring ordering in rapidity. $\alpha_0$ is some initial
value of $\alpha$ at which the initial conditions are set. In the
first step of the evolution in \eq{rintev} the ordering in rapidity is
imposed by using both distances $|\un z - \un x|$ and $|\un z - \un
y|$ as cutoffs, since the hard quark is the one we are tagging on and
it has two different transverse coordinates on both sides of the
cut. The subsequent steps of the evolution are easier, since there the
harder quark will only have one coordinate. The resulting quantity
$r_{int} ({\un w}, \un w; {\un z}, \alpha)$ obeys the following
evolution equation, which is obtained from \eq{rintev} by simply
putting $\un x = \un y = \un w$:
\beq\label{rintev1}
r_{int} (\un w, \un w ; \un z, \alpha) \, = \, r_{int}^{(0)} (\un z, \alpha) \, + \,
\frac{\bas}{2 \, \pi} \, \int_{\alpha_0}^{\alpha \,
\mbox{min} \left\{1, \frac{(\un z - \un w)^2}{(\un z - \un z_2)^2} \right\}} \, 
\frac{d \alpha'}{\alpha} \, \frac{d^2 z_2}{(\un z - \un z_{2})^2} \, 
r_{int} ({\un z}, \un z;  {\un z}_2, \alpha'). 
\eeq

With the help of Eqs. (\ref{sig_hard}), (\ref{brepl}) and
(\ref{rintev}) we write the expression for the valence quark
production cross section with the quark originating in the nuclear
wave function including the effects of small-$x$ evolution
\ben
\frac{d\sigma^b}{d^2 k\,dy} (\un b_{01}) \, = \, \frac{1}{2 \, (2 \pi)^4} \, 
\int d^2 x \, d^2 y \, d^2 z \, d^2 B \, d^2 b_{0'} \, d^2 b_{1'} \, 
n_1 ({\un b}_{0}, \un b_1; {\un b}_{0'}, \un b_{1'}; Y-y) \, e^{i \un
k \cdot (\un x - \un y)} 
\een
\beq\label{bev}
\times \, \int_0^1 d \alpha \, 
\sum_{i,j=0}^1 \, \Phi (\un x - \un b_{i'}, \un y - \un b_{j'}; \un z, \alpha) \,
r_{int} (\un x, \un y ; \un z, \alpha) \, [1 - N (\un x, \un y, y)].
\eeq

Integrating \eq{bev} over $k_T$ gives the following expression for the
rapidity distribution of the stopped baryon number
\ben
\frac{d\sigma^b}{dy} (\un b_{01}) \, = \, \frac{1}{2 \, (2 \pi)^2} \, 
\int d^2 x \, d^2 z \, d^2 B \, d^2 b_{0'} \, d^2 b_{1'} \, 
n_1 ({\un b}_{0}, \un b_1; {\un b}_{0'}, \un b_{1'}; Y-y)
\een
\beq\label{bevy}
\times \, \int_0^1 d \alpha \, 
\sum_{i,j=0}^1 \, \Phi (\un x - \un b_{i'}, \un x - \un b_{j'}; \un z, \alpha) \,
r_{int} (\un x, \un x ; \un z, \alpha).
\eeq
It is interesting to note that, as could be expected, all the
non-linear effects disappeared from $d\sigma^b/dy$. 

\begin{figure}[th]
\begin{center}
\includegraphics[width=14cm]{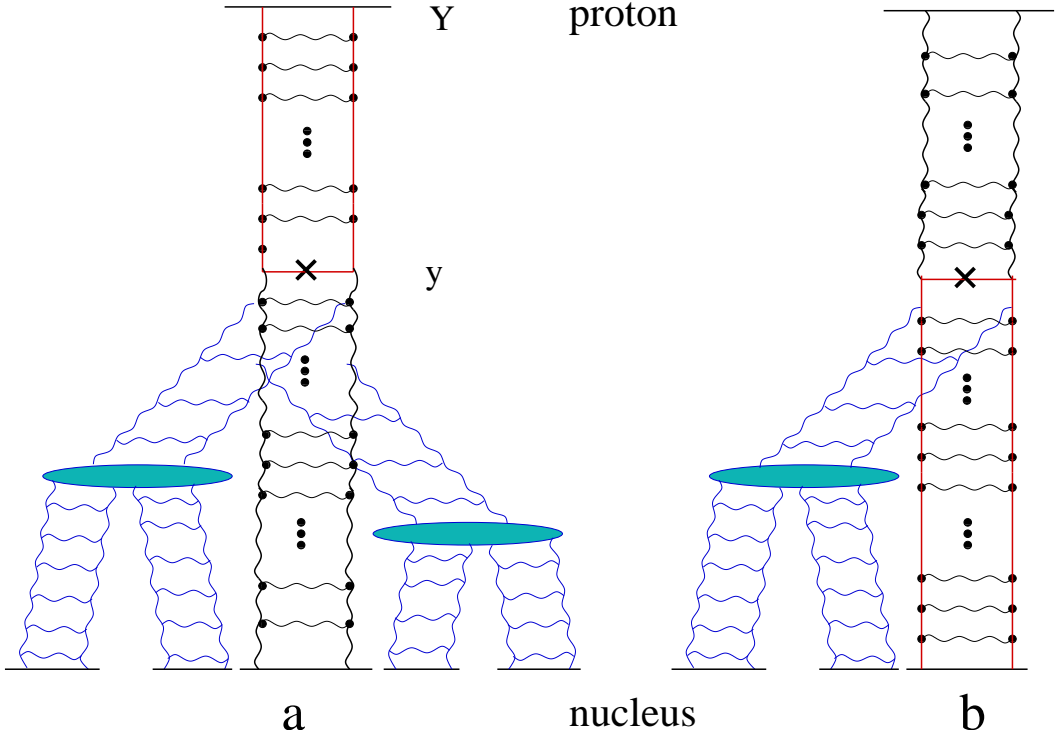}
\end{center}
\caption{Fan diagrams corresponding to the 'a' and 'b' contributions to the
  valence quark production cross section including the quantum
  evolution effects given by Eqs. (\ref{dsaf5}) and (\ref{bev})
  correspondingly.}
\label{fan}
\end{figure}

The essential ingredients of \eq{bev} are the Reggeon exchange
amplitude $r_{int}$ obeying the linear equation (\ref{rintev}) summing
up powers of $\as \, \ln^2 s$, and the amplitudes $n_1$ and $N$
obeying Eqs. (\ref{eqn}) and (\ref{eqN}) correspondingly, which sum
powers of $\as \, \ln s$. Note that at high $k_T$, in the regime where
evolution is linear, one can neglect $N$ in \eq{bev}, and, keeping
only the linear (BFKL) part of \eq{eqN} for $N$ in \eq{dsaf5}, one can
show that each of the equations (\ref{dsaf5}) and (\ref{bev}) become
dependent on a convolution of the BFKL pomeron and Reggeon exchange
amplitudes.

The fan diagram representation of the valence quark production cross
sections calculated in Eqs. (\ref{dsaf5}) and (\ref{bev}) are shown in
\fig{fan}. The diagram 'a' in \fig{fan} illustrates a contribution to
\eq{dsaf5} with the valence quark originating in the proton: the
diagram has a linear Reggeon evolution between the proton and the
produced valence quark, with a full non-linear evolution of \eq{eqN}
represented by fan diagrams between the produced quark and the target
nucleus. The diagram 'b' in \fig{fan} displays a contribution to
\eq{bev} with the valence quark originating in the nucleus: it has a
linear BFKL evolution between the proton and the produced valence
quark, a linear Reggeon evolution between the produced quark and the
target, along with a possibility of a non-linear splitting attached to
the produced valence quark line.


\section{Conclusions}

Above we presented calculations of valence quark production far from
the fragmentation region in $pA$ collisions. The answers for
production cross section in the quasi-classical approximation are
given by Eqs. (\ref{dsaf}) and (\ref{dsbf}) corresponding to the
valence quark originating in the proton and nucleus. The
quasi-classical results lead to Cronin-like enhancement of nuclear
modification factor for stopped baryons, as shown in
\fig{cronin}. Obtained transverse momentum spectra of the produced
valence quarks scale as $\sim 1/k_T^3$ at high $k_T$ and at fixed
rapidity, indicating higher sensitivity of quark production to the
physics at the ultraviolet end of the spectrum.

Quantum evolution corrections, in the sense of double logarithmic
reggeon evolution of \cite{KL,Kirschner1,Kirschner2,GR} and single
logarithmic gluon evolution of \cite{yuri,bal} are included in the
valence quark production cross section with the answers given by
Eqs. (\ref{dsaf5}) and (\ref{bev}) corresponding to the valence quark
originating in the proton and in the nucleus. \eq{dsaf5} contains the
effects of double-logarithmic linear reggeon evolution in the rapidity
interval between the produced quark and the projectile along with
non-linear gluon evolution effects between the quark and the target
nucleus. The nuclear modification factor of stopped baryons coming
from the projectile proton (due to
\eq{dsaf5}) is explored and shown to give suppression of the produced
baryons at higher energies and/or rapidities (see \fig{evcron}). 
Therefore, based on the valence quark production spectrum and
barring the effects of fragmentation functions and of non-perturbative
mechanisms of baryon stopping \cite{Veneziano,RV,KZ,Dima96,CS}, we
expect the stopped baryon spectrum to exhibit suppression in the
forward rapidity region in $d+Au$ collisions at RHIC. Indeed similar
effects should be expected at LHC (possibly even at mid-rapidity
\cite{KKT2}) if a $pA$ run is to be conducted there.

\eq{bev} gives us the baryon stopping cross section with the valence 
quark coming from the nucleus. It includes linear BFKL evolution
between the produced quark and the projectile and a combination of
linear reggeon evolution and the non-linear evolution from \eq{eqN} in
the rapidity interval between the quark and the target. The study of
the impact of small-$x$ reggeon evolution in \eq{bev} on nuclear
modification factor is left for future work. Qualitatively we expect
weaker suppression than for baryons coming from the proton, but the
onset of this suppression should happen at smaller rapidities. Indeed
for the non-linear evolution effects to become important the valence
quark should be sufficiently far away in rapidity from the target. Due
to the double-logarithmic nature of reggeon evolution one should
expect it to manifest itself at smaller rapidities than the gluon
evolution. At the same time, as the valence quark production cross
section in \eq{bev} falls off approximately as $\sim e^{-y}$, the
non-linear evolution effects in it are likely to become important in
the rapidity region where baryon stopping in the nucleus (i.e., the
cross section itself) is small. This is likely to be the case at RHIC,
where the small-$x$ evolution effects do not become important until
forward rapidity: there baryon stopping will be dominated by the
valence quarks coming from the projectile deuteron (or proton) and
baryon stopping due to \eq{dsaf5} will prevail over that from
\eq{bev}. In other words evolution effects in baryon stopping in the
nucleus for $d+A$ collisions at RHIC may become important only in the
rapidity region where baryon stopping is dominated by baryons from the
deuteron.  Therefor, in order to test \eq{bev} one has to perform $pA$
collision experiments at the LHC, where even at mid-rapidity one would
expect significant small-$x$ evolution effects \cite{KKT2}. There one
may find a window in rapidity (probably close to mid-rapidity but also
not too distant from the nuclear fragmentation region) where baryon
stopping is dominated by the mechanism given by \eq{bev} with the
non-linear evolution effects being important.

The results of our calculations are generally applicable to $pA$
collisions to quantify the amount of purely perturbative baryon
stopping. In the high-$p_T$ sector, for $p_T > Q_s$, our results are
also applicable to nucleus-nucleus collisions. It is likely, though by
no means proven, that the obtained cross sections would provide a
qualitatively correct description of the nuclear modification factor
for the perturbative initial-state baryon stopping in nucleus-nucleus
collisions even at $p_T < Q_s$. Indeed at rapidity close to the
fragmentation region of one of the nuclei in $A+A$ collisions one may
argue that the saturation scale of that nucleus ($Q_{s2}$) would be
small, much smaller than the saturation scale of the other nucleus
($Q_{s1}$), making the particle production processes describable in
the framework of $pA$ collisions for a wide range of transverse
momenta $p_T > Q_{s2}$. Therefore in that rapidity region baryon
stopping can be described by the \eq{dsaf5} above. The amount of
baryon stopping not far from the fragmentation region is large,
allowing for an easy experimental verification of our results.

We conclude by reiterating that suppression of high-$p_T$ net baryons
in $p(d)+A$ collisions at forward rapidity at RHIC, as predicted
above, would provide an independent test of the physics of Color Glass
Condensate. (Non-perturbative baryon stopping mechanisms are not
likely to be important at high-$p_T$.) Similar experiments can be
carried out at the LHC. The proposed test of CGC at RHIC with valence
quarks may not be as clean as electromagnetic probes \cite{emprobes},
which are free from uncertainties introduced by fragmentation
functions, but it has the advantage of being possible to perform by
analyzing the current experimental data generated at RHIC.

\section*{Acknowledgments} 

This work is supported in part by the U.S. Department of Energy under
Grant No. DE-FG02-05ER41377.


\renewcommand{\theequation}{A\arabic{equation}}
  \setcounter{equation}{0} 
  \section*{Appendix A}

\begin{figure}[ht]
\begin{center}
\includegraphics[height=2.8cm,width=5.5cm]{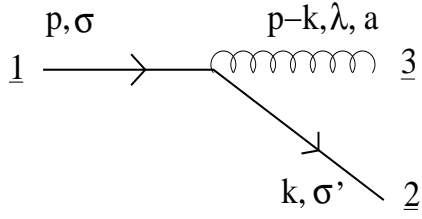}
\end{center}
\caption{Diagram for soft quark wavefunction calculation.}
\label{qwf_fig}
\end{figure}

In this Appendix we derive the small-x quark and gluon wavefunction of
a fast quark moving in the light cone ``plus'' direction. The
corresponding diagram is shown in \fig{qwf_fig}. At lowest order in
perturbation theory it is given by \cite{BL}
\beq
\psi^{'a}_{\sigma,\sigma'\lambda}(\ud{p},\ud{k},\alpha)=
\frac{\langle f|H_I|i\rangle/\sqrt{k^+p^+}}{\sum_fp_f^--\sum_ip_i^-}, 
\eeq
\noi where $p$ and $\sigma$ are the momentum and helicity of the incoming
quark, $k$ and $\sigma'$ those of outgoing quark and $a$ and $\lambda$ 
the color and polarization of the emitted gluon. The transition matrix element
is given by
\beq
\langle f|H_I|i\rangle/\sqrt{p^+k^+} \, = \, g \, t^a \frac{\bar{u}_{\sigma}(k)}{
\sqrt{k^+}}\gamma\cdot\epsilon^{\lambda}(p-k) \frac{u_{\sigma'}(p)}{\sqrt{p^+}},
\eeq
\noi where the gluon polarization vector in in the $A^+=0$ gauge is\footnote{Here we
  follow the convention in which the product of two arbitrary vectors
  is given by $a\cdot b=\frac{1}{2}(a^+b^-+a^-b^+)-\ud{a} \, \cdot \,
  \ub$} $\epsilon(q)=(0,\frac{2q\cdot\epsilon^{\lambda}}{
  q^+},\ud{\epsilon}^{\lambda})$ with $\ud{\epsilon}^{\lambda}=
  \frac{1}{\sqrt{2}}(1,i\lambda)$ and $\lambda=\pm1$.

\noi Defining $\alpha=k^+/p^+$, the energy denominator is
\beq
\Sigma_fp_f^--\Sigma_ip_i^-=-\frac{1}{p^+\alpha(1-\alpha)}(\alpha\ud{p}-\uk)^2
\eeq

\noi We note that \cite{BL}
\beq
\frac{\bar{u}_{\sigma}(k)}{
\sqrt{k^+}}\gamma^+ \, \frac{u_{\sigma'}(p)}{\sqrt{p^+}}=2\delta^{\sigma\sigma'}
\eeq
and 
\beq
\frac{\bar{u}_{\sigma}(k)}{
\sqrt{k^+}}\gamma^j \frac{u_{\sigma'}(p)}{\sqrt{p^+}}=\left[\frac{k^j-i\sigma
\epsilon^{jl}k^l}{ k^+} + \frac{p^j+i\sigma\epsilon^{jl}p^l}{p^+}\right]
\delta^{\sigma\sigma'} 
\eeq
\noi where $\epsilon^{jl}$ the totally antisymmetric tensor in two
dimensions. Using $i\sigma\uk^l\epsilon^{jl}\ud{\epsilon}^{\lambda}_j=
\sigma\lambda\ud{k}\cdot\ud{\epsilon}^{\lambda}$ one immediately gets
\beq
\psi^{'a}_{\sigma,\sigma'\lambda}(\ud{p},\ud{k},\alpha)=gt^a[1+\alpha-(1-\alpha)
\sigma\lambda]
\frac{(\uk-x\ud{p})\cdot\ud{\epsilon}^{\lambda}}{(\uk-x\ud{p})^2}
\delta^{\sigma\sigma'}.
\eeq

\noi Denoting by $\ux_1$ the transverse coordinate of the
incoming quark and by $\ux_2$ and $\ux_3$ the transverse coordinates
of the outgoing quark and gluon respectively and Fourier-transforming
to coordinate space one gets
\beq
\psi^{'a}_{\sigma,\sigma'\lambda}(\ux_{31},\ux_{23},\alpha)
=gt^a[1+\alpha-(1-\alpha)\sigma\lambda] \frac{i}{2\pi}
\delta(\ux_{31}+\alpha\,\ux_{23})
\frac{\ud{x}_{23}\cdot\ud{\epsilon}^{\lambda}}{\ux^2_{23}}
\delta^{\sigma\sigma'}.
\label{wf}
\eeq

\noi Taking the limit $\alpha\rightarrow 0$ and integrating over the
transverse position of the hard gluon sets $\ux_3=\ux_1$ due to the 
delta-function and yields
\beq\label{small-x}
\psi^a_{\sigma,\sigma'\lambda}(\ux_{21},\alpha)
=\int d^2x_3 \psi^{'a}_{\sigma,\sigma'\lambda}(\ux_{31},\ux_{23},\alpha)
=gt^a[1-\sigma\lambda]\frac{i}{2\pi}
\frac{\ud{x}_{21}\cdot\ud{\epsilon}^{\lambda}}{\ux^2_{21}}
\delta^{\sigma\sigma'}.
\eeq

\noi To define the small-$\alpha$ quark distribution function of a nucleon we
multiply the previous wavefunction by its complex conjugate evaluated
at different transverse position of the soft quark, average over the
helicity of the initial quark and sum over the polarizations and
colors of the gluon and integrate over the position of the initial
quark ($\ux_1$) obtaining
\beq
\frac{dn^q}{d^2z \, dy}=\frac{1}{4\pi} \frac{\alpha}{2}\sum_{\sigma,\lambda,a}
\langle\psi^a_{\sigma,\lambda}(\ux-\uz,\alpha)
\psi^{*a}_{\sigma,\lambda}(\uy-\uz,\alpha) 
\rangle =\frac{\bar{\alpha}_s\,\alpha}{2\pi} \frac{(\ux-\uz)\cdot(\uy-\uz)
}{(\ux-\uz)^2(\uy-\uz)^2},
\eeq
as given in \eq{qwf}. The factor $\alpha/2$ arises from the integration
over the phase space of the gluon: $d\alpha/2(1-\alpha)\approx \alpha dy/2$.

\noi Taking the limit $1-\alpha\rightarrow 0$ in \eq{wf} and proceeding
analogously one derives the soft gluon distribution function of a
valence quark
\beq
A^a_{\sigma,\sigma'\lambda}(\ux_{31},\alpha)
= \, 2 \, g \, t^a \, \frac{i}{2\pi}
\frac{\ud{x}_{31}\cdot\ud{\epsilon}^{\lambda}}{\ux^2_{31}}
\delta^{\sigma\sigma'}\,
\eeq
and 
\beq
\frac{dn^g}{d^2z \, dy}=\frac{1}{4\pi} \, \frac{1}{2} \, 
\sum_{\sigma,\lambda,a}
\langle A^a_{\sigma,\lambda}(\ux-\uz,\alpha)A^{*a}_{\sigma,\lambda}
(\uy-\uz,\alpha)
\rangle = \frac{\bar{\alpha}_s}{\pi} \frac{(\ux-\uz)\cdot(\uy-\uz)
}{(\ux-\uz)^2(\uy-\uz)^2}.
\eeq

\renewcommand{\theequation}{B\arabic{equation}}
  \setcounter{equation}{0} 
  \section*{Appendix B}
\begin{figure}
\begin{center}
\includegraphics[height=4cm,width=5cm]{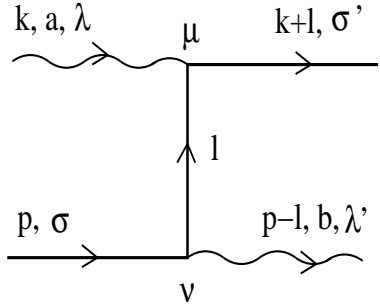}
\caption{Gq$\rightarrow$qG scattering amplitude.}
\end{center}
\label{gqcs}
\end{figure}

\noi In this appendix we derive the $Gq\rightarrow qG$ scattering cross 
section in the high energy limit. The calculation is performed in the 
$A^+=0$ gauge, with gluon polarization vectors taken in the same gauge
$\epsilon^+=0$. The amplitude for the process is shown in \fig{gqcs}
and is equal to
\beq
M=\bar{u}_{\sigma'}(k+l)(igt^a\gamma^{\mu}) \frac{i\slashed{l}}{l^2+i\epsilon}
(igt^b\gamma^{\nu})u_{\sigma}(p)\epsilon_{\mu}^{\lambda}(k)
\epsilon_{\nu}^{*\lambda'}(p-l).
\eeq
\noi Squaring the amplitude, averaging over the helicity of the incoming quark
and the polarization of the incoming gluon and summing for all possible
helicities, polarizations and colors in the final state, one gets 
\bem
\left\langle|M|^2\right\rangle \, \equiv \, \frac{1}{4} \, \frac{1}{
  N(N^2-1)}\sum_{a, b, \sigma, \sigma', \lambda, \lambda'}|M|^2 \\
=(2\pi)^2 \frac{C_F}{2N_c} \frac{\alpha_s^2}{(l^2+i\epsilon)^2}\mbox{tr}[(\s{k}+\s{l})
\gamma^{\mu}\s{l}\gamma^{\nu}\s{p}\gamma^{\beta}\s{l}\gamma^{\alpha}]
d_{\mu\alpha}(k)d_{\nu\beta}(p-l)
\end{multline}
\noi where $d_{\mu\alpha}(k)=\sum_{\lambda}\epsilon_{\mu}^{\lambda}(k)
\epsilon_{\alpha}^{*\lambda}(k)= -g_{\mu\alpha}+\frac{k_{\alpha}\eta_{\mu}+
  k_{\mu}\eta_{\alpha}}{ k\cdot\eta}$ is the gluon polarization sum in
  the light cone gauge.

\noi Keeping only the leading terms, i.e. those proportional to the 
center mass energy of the collision, $s=2p^-k^+$, and putting 
$l^2\approx -\ud{l}^2$, we get
\beq
\left\langle|M|^2\right\rangle=4(2\pi)^2 \frac{\alpha_s^2C_F}{N_c}
\frac{s}{\ud{l}^2}\;.
\eeq

\noi Finally, dividing by the flux factor and integrating over the 
phase space of the produced quark and gluon one gets the final result
\beq
\sigma^{Gq\rightarrow qG} \, = \, \frac{\alpha_s^2C_F}{N_c} \, 
\frac{1}{s}\int\! 
\frac{d^2\ud{l}}{\ud{l}^2}\;.
\eeq


\end{document}